\pdfoutput=1
\documentclass[
  aps,
  prd,
  superscriptaddress,
  nofootinbib,
  amsmath,
  amssymb,
  longbibliography,
  twocolumn
]{revtex4-2}
\usepackage{physics}
\usepackage{mathtools}
\usepackage{multirow}

\usepackage{hyperref}
\usepackage{orcidlink}
\hypersetup{
    colorlinks=true,       
    linkcolor=red,          
    citecolor=blue,        
    filecolor=magenta,     
    urlcolor=cyan          
}

\usepackage{newtxtext}
\usepackage{newtxmath}

\begin{document}
\title{Universal Structure of Horizon Formation in Generic Binary Black Hole Mergers}
\author{Yu-Cun Xie\,\orcidlink{0000-0002-7603-7411}}
\email{yzx5296@psu.edu}
\affiliation{Institute for Gravitation and the Cosmos, The Pennsylvania State University, University Park, PA 16802, USA}
\affiliation{Department of Physics, The Pennsylvania State University, University Park, PA 16802, USA}
\author{Vaishak Prasad\,\orcidlink{0000-0001-6712-2457}}
\affiliation{Institute for Gravitation and the Cosmos, The Pennsylvania State University, University Park, PA 16802, USA}
\affiliation{Department of Physics, The Pennsylvania State University, University Park, PA 16802, USA}
\date{November 11, 2026}
\begin{abstract}
A new common apparent horizon forms during a binary black hole merger and gives rise to outer and inner branches of marginally outer trapped surfaces. We derive the local structure of these branches.  This event occurs in the fully nonlinear regime,
outside the standard regimes of post-Newtonian inspiral theory and perturbations of a stationary black hole,
yet it admits a universal description.  Without assuming symmetry, we show that the stability operator of
the marginally outer trapped surface must lose invertibility at formation.  Outermost stability then implies that the vanishing eigenvalue is
the principal one, with a strictly positive eigenfunction.  Lyapunov--Schmidt reduction yields square-root branch separation with a shared linear drift. The outer and inner common horizons lie on a smooth marginally outer trapped tube along which slice time has a local minimum at formation, and nearby later slices intersect it in outer and inner branches whose separation scales as $(t-t_*)^{1/2}$.  Sufficiently smooth quasilocal scalar functionals with nonzero first variation along the principal zero mode inherit the square-root separation and a shared linear term.  We test these
predictions in three binary black hole simulations, including an eccentric,
precessing, unequal-mass system.  In all three, the
worldtube geometry and quasilocal scalars follow the predicted scaling.  With the next-order term included, free-exponent fits to the surface geometry and quasilocal functionals recover the square-root exponent, $1/2$, to within half a percent, and diagnostics agree on the formation time within $5\times10^{-5}M$. The correlation between horizon shear and gravitational-wave news raises the possibility that common horizon formation has a signature in a short segment of the merger waveform.  
\end{abstract}
\maketitle
\newpage

\section{Introduction}\label{sec:intro}
The numerous direct detections of gravitational waves from compact-binary coalescence have transformed the study of the general relativistic two-body problem into an observationally driven field.  Binary black-hole mergers are especially important in this context because the late inspiral, merger, and ringdown probe a regime in which velocities are relativistic, spacetime curvature is large, and the Einstein equations are fully nonlinear.  The first detection \cite{LIGOScientific:2016aoc} provided evidence for a binary black hole merger whose waveform was consistent with the predictions of general relativity, and subsequent tests \cite{LIGOScientificVirgo:2016gri} used the same event to probe the dynamical strong-field regime of the theory. More recently, the high signal-to-noise ratio detections GW230814 \cite{LVK:2026GW230814}, GW231123 \cite{LIGOScientificVirgoKAGRA:2025GW231123}, and GW250114 \cite{LIGOScientific:2025rid} have enabled more detailed studies of binary black hole merger and ringdown signals.  Gravitational-wave astronomy, therefore, motivates not only accurate waveform modeling \cite{HealyLousto:2022RIT,Boyle:2019SXS,Ferguson:2025MAYA} but also a more detailed understanding of the spacetime geometry that produces the radiation.

A central notion in any black hole spacetime is the horizon.  In the global description, a black hole is defined through its event horizon, the boundary of the causal past of future null infinity.  This definition is  teleological as defining an event horizon requires knowledge of the entire future development of the spacetime.  In numerical relativity, event horizons are therefore reconstructed only after the spacetime evolution is available, typically by propagating null generators backward from a sufficiently late time \cite{Libson:1996nt,Cohen:2008yb}.  Moreover, the topology of their spatial cross-sections during merger depends on the choice of slicing \cite{Cohen:2011nk,Bohn:2016afc}.  Thus an event horizon does not provide a local-in-time marker of black hole formation on a given foliation \cite{Thornburg:2007eventAHfinders}.  Numerical relativity, therefore, relies on quasilocal notions of black-hole boundaries.  Following Hawking, an apparent horizon is the outer boundary of the trapped region on a spatial slice \cite{Hawking:1973eventHorizon,HawkingEllis1973}.  This boundary can be characterized as an outermost marginally outer trapped surface \cite{KrieleHayward:1997,AnderssonMetzger:2009}.  These surfaces can be located on individual Cauchy slices and give a practical way of identifying and characterizing black holes during a simulation \cite{Thornburg:2003sf,Thornburg:2007eventAHfinders,Etienne:2026BHaHAHA}.

The notion of a trapped surface, introduced by Penrose, provides a quasilocal way to identify strong gravitational focusing.  A smooth, closed, spacelike two-surface is future trapped when the expansions of both independent future-directed null normals are negative.  Physically, both the ingoing and the outgoing families of orthogonal light rays are locally converging, so even light directed outward fails to expand away from the surface.  This condition is determined by the geometry at the surface.  Under suitable energy, causality, and global assumptions, the existence of a closed trapped surface implies future null geodesic incompleteness, forming the basis of the Penrose and Hawking--Penrose singularity theorems \cite{Penrose1965,HawkingPenrose1970,HawkingEllis1973}.  
A marginally outer trapped surface (MOTS) is a closed surface whose outgoing null expansion vanishes.
A smoothly continued family of MOTSs traces a three-dimensional hypersurface called a marginally outer trapped tube (MOTT), a quasilocal horizon (for reviews, see Refs.~ \cite{Booth:2005boundaries,Krishnan:2008quasilocal,AshtekarKrishnan:2025recent}).  

The modern theory of quasilocal horizons was developed through the frameworks of trapping horizons, non-expanding horizons, isolated horizons, and dynamical horizons \cite{Hayward:1993wb,Ashtekar:1998sp,Ashtekar:2004cn}.  Hayward's trapping-horizon framework classifies future and past as well as outer and inner horizons and yields local area laws \cite{Hayward:1993wb}.  A non-expanding horizon is a null quasilocal horizon whose intrinsic geometry and cross-sectional area do not change along its null generators.  An isolated horizon additionally requires the intrinsic connection to be time independent, while allowing the exterior spacetime to remain dynamical \cite{Ashtekar:1998sp,Ashtekar:2004cn}.  Dynamical horizons describe growing black holes and obey flux balance laws for area, mass, and angular momentum derived by Ashtekar and Krishnan \cite{AshtekarKrishnan:2002flux,AshtekarKrishnan:2003properties}. Hayward later developed related balance laws in a dual-null formulation \cite{Hayward:2004fz}.  Together these frameworks support quasilocal definitions of area, mass, angular momentum, shear, and fluxes directly on the black hole boundary \cite{Ashtekar:2004cn,AshtekarEnglePawlowski:2004,AshtekarKrishnan:2025recent}.  Recent work extends the thermodynamic description to strongly dynamical regimes \cite{Ashtekar:2025qqa,Ashtekar:2026jdz}.  Applications include black hole entropy in quantum gravity, horizon evolution in numerical relativity, and comparisons with gravitational-wave signals.

Quasilocal horizons also encode information about the strong-field source of the radiation.  Numerical studies have found that the shear of outgoing null rays at the horizons correlates with the gravitational-wave news, both on the individual horizons during inspiral \cite{Prasad:2020news} and on the outer common horizon after formation \cite{Prasad:2025shear}. Related studies have examined tidal distortions of the horizons and their relation to the emitted waveform \cite{PrasadGuptaBoseKrishnan:2022tidal,Prasad:2024tidalImprint}.  Other work has characterized the quasinormal mode structure of the common horizon \cite{Mourier:2021commonHorizonQNMs,Chen:2022commonHorizonMultipoles,Prasad:2025shear}, including quadratic contributions to the shear of the outer common horizon following head-on mergers \cite{Khera:2023nonlinearRingdownHorizon}.  Horizon tomography makes this connection quantitative in a perturbative ringdown setting by relating radiative fields at the horizon and at null infinity \cite{RibesMetidieriBongaKrishnan:2025tomography}. These studies suggest that gravitational-wave observations could constrain horizon geometry.

\begin{figure}[htbp]
    \centering
    \includegraphics[width=\columnwidth]{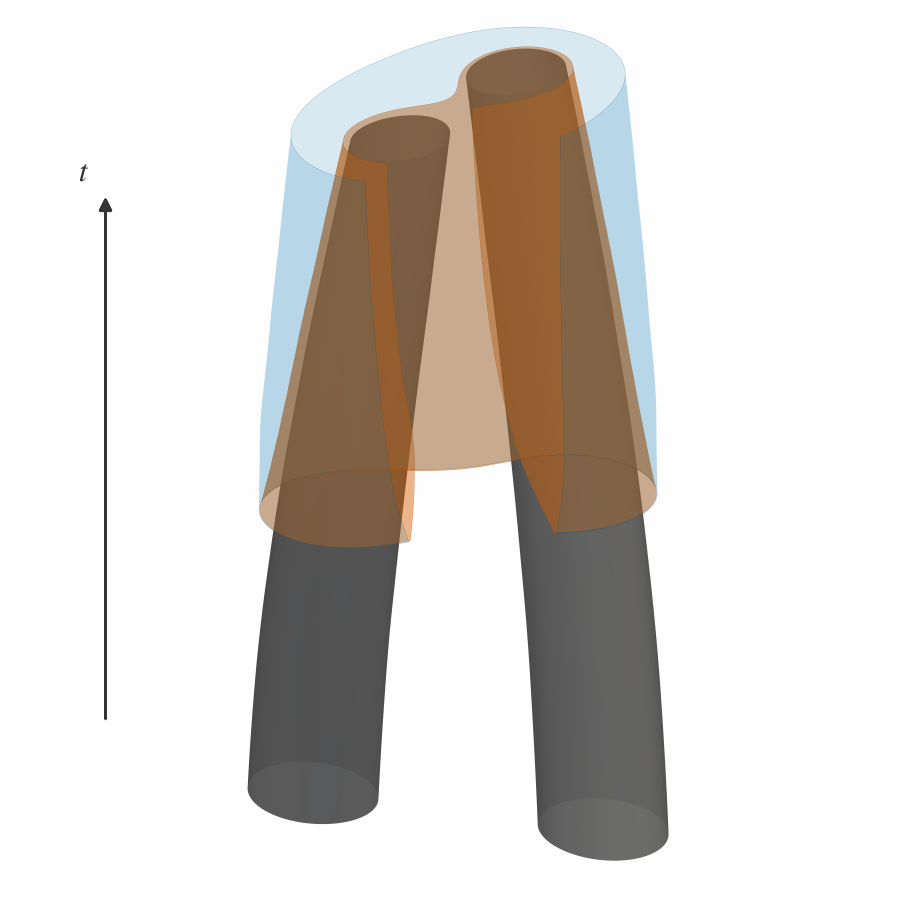}
    \caption{Formation of the common horizon in a binary black hole merger (Simulation~I of Sec.~\ref{sec:numerical}), with one non-equatorial dimension suppressed. The common horizon appears as a single two-sphere at some $t_*$ and then bifurcates into two branches, an outer branch with increasing area (blue) and an inner branch with decreasing area (vermilion). Both common branches are cut open to show the two individual horizons (gray) nested inside.}
    \label{fig:ill}
\end{figure}

Within a chosen foliation, the first common apparent horizon marks the formation of a single remnant black hole.  The common apparent horizon does not form by joining the individual apparent horizons at contact.  Instead, numerical evolutions show that a new common apparent horizon appears outside the two individual horizons once they are sufficiently close \cite{Gupta2018DynamicsMarginallyTrappedSurfaces,Pook-Kolb:2018igu,Pook-Kolb:2019iao}, as illustrated in Fig.~\ref{fig:ill}.  The two-black-hole apparent-horizon problem was studied in time-symmetric initial data by \v{C}ade\v{z} and others \cite{Cadez:1974,Bishop:1982closedTrappedRegion,CookAbrahams:1992horizonStructure}, and early numerical evolutions explicitly observed the appearance of a larger common horizon around the two colliding holes \cite{Anninos:1994apparent}. For perturbative apparent-horizon calculations in Brill--Lindquist data, see Ref.~\cite{Krishnan:2002thesis}.  This behavior is deeply tied to the geometry of trapped surfaces.  Andersson, Mars, Metzger, and Simon showed \cite{Andersson:2008up} that, under suitable hypotheses, when the two individual components of the outermost MOTS approach one another, the outermost MOTS must jump before the components touch; the boundary of the resulting connected trapped region is a new common MOTS enclosing the progenitors.  Thus, within a chosen foliation, the birth of the common apparent horizon is a property of the trapped region rather than a numerical artifact.

The local behavior of MOTSs is controlled by the MOTS stability operator, the linearization of the outgoing null expansion under deformations of the surface.  This operator is elliptic \cite{Andersson:2007fh}. Its principal eigenvalue is real and controls outward stability. Andersson, Mars, and Simon established its spectral properties and showed that strict stability gives local existence of a MOTT \cite{Andersson:2005gq,Andersson:2007fh}. When the principal eigenvalue vanishes, the tube can become tangent to a Cauchy slice \cite{Andersson:2008up}, as expected at first common horizon formation \cite{Pook-Kolb:2018igu}.

Numerical studies over the last several years have revealed that the horizon structure inside a binary black-hole merger is far richer than the appearance of a single common apparent horizon enclosing the two progenitors \cite{Gupta2018DynamicsMarginallyTrappedSurfaces,Pook-Kolb:2018igu,Pook-Kolb:2020zhm,Pook-Kolb:2020jlr}.  Methods have been developed that can track highly distorted and self-intersecting MOTSs assuming axisymmetry, showing that the interior of a merger contains multiple MOTSs, including unstable MOTSs that bifurcate and annihilate \cite{Pook-Kolb:2019iao,Pook-Kolb:2019ssg}.  Subsequent work showed that MOTS worldtubes can weave back and forth in time \cite{Pook-Kolb:2021gsh,Booth:2021sow,Pook-Kolb:2021jpd}. Related studies have examined self-intersecting MOTSs in Schwarzschild spacetime \cite{BoothHennigarMondal:2020MOTS} and extended MOTS-finding methods to rotating, axisymmetric spacetimes \cite{BoothEtAl:2023rotatingMOTS}.  These results strongly suggest that the spectrum of the stability operator provides a natural way to understand the bifurcation of horizons during merger.

We seek a quantitative analytic description of common horizon formation within this fully nonlinear regime. Existing stability results identify a vanishing principal eigenvalue and the tangency of a MOTT to the formation slice \cite{Andersson:2008up,Pook-Kolb:2018igu}. Building on this structure, we determine the local branch expansion through linear order in time, with explicit expressions for the coefficients of both branch separation and shared drift, and derive the corresponding laws for horizon quantities.

We consider a reference MOTS at the first common-horizon formation slice and represent nearby trial surfaces by a normal graph height over it, meaning the displacement along the fixed reference normal.  The MOTS equation is then expanded in the graph height and the time displacement of the slice.  Because no common MOTS exists on nearby earlier slices, an invertible linearized operator would incorrectly imply a smooth family extending to those slices.  The linearized operator must therefore have a zero mode $H_0$, satisfying $L_\Sigma[H_0]=0$, at first common-horizon formation, and outermost stability identifies it with a vanishing principal eigenvalue. Lyapunov--Schmidt reduction determines the amplitude of the zero mode through a scalar equation and the remaining surface deformation through a complementary equation.  The resulting equation can be solved order by order and gives two nearby horizon branches whose separation scales as the square root of the time from formation. The leading-order reduced equation has the normal form of a fold, or saddle-node, bifurcation \cite{strogatz2014nonlinear}, and the concurrent work of Booth, Cox, and Okpala studies more general bifurcations in black hole horizon evolution \cite{BoothCoxOkpala:2026bifurcation}. The subsequent analysis gives a scaling law for a sufficiently regular quasilocal scalar functional $\mathcal Q[H,t]$ evaluated on the two horizon branches, provided its first variation along the direction of the zero mode is nonzero,
\begin{equation}
        \mathcal{Q}_{\pm}(t)-\mathcal{Q}_* =
    \pm \mathcal{A}_{\mathcal{Q}}\,(t-t_*)^{1/2}
    + \mathcal{B}_{\mathcal{Q}}\,(t-t_*)
    + O\!\left((t-t_*)^{3/2}\right).
\end{equation}
Here $\mathcal Q_\pm(t)$ are the functional's values on the two branches and $\mathcal Q_*$ is its value at formation. Through the retained order, the local branch structure is a tilted parabola in the \((t,\mathcal Q)\) plane: the branches separate as $(t-t_*)^{1/2}$ and share a drift proportional to $t-t_*$.  The leading-order exponent is fixed, while the coefficients are computable from geometric data on the formation surface and the variations of the MOTS equation.  Under the stated assumptions, this gives a generic prediction for the short-time evolution of horizon quantities near common horizon formation.

The paper is organized as follows.  Section~\ref{sec:basic} reviews basic concepts related to MOTSs, MOTTs, the representation of nearby surfaces and the stability operator.  Section~\ref{sec:theory} derives the local branch equation at common horizon formation and the resulting square-root scaling law for quasilocal quantities.  Section~\ref{sec:numerical} tests the predictions in three binary black hole simulations, extracts the opening and drift fields, and uses the scaling law to recover the formation time from later converged horizons.  Section~\ref{sec:discussion} interprets the formation fields and discusses extensions and applications.

\section{Basic notions}\label{sec:basic}

\subsection{Quasilocal horizons and marginally outer trapped surfaces}

Consider a spacetime foliated by spacelike Cauchy surfaces \(\Sigma_t\).  The induced metric on \(\Sigma_t\) is denoted by \(\gamma_{ij}(t)\), and its extrinsic curvature by \(K_{ij}(t)\), with trace \(K=\gamma^{ij}K_{ij}\).  Let \(S\subset\Sigma_t\) be a smooth closed two-surface with outward-pointing spatial unit normal \(s^i\).  If \(n^a\) is the future-directed unit normal to \(\Sigma_t\), we fix the normalization of the outgoing and ingoing null normals by choosing
\begin{eqnarray}
    \ell^a = \frac{n^a+s^a}{\sqrt{2}}, \qquad
    k^a = \frac{n^a-s^a}{\sqrt{2}} .
\end{eqnarray}
With this convention, \(\ell^a k_a=-1\).

The outgoing null expansion is the fractional rate of change of the area element of \(S\) along the outgoing null congruence.  With the sign conventions used here,
\begin{equation}
        \Theta_{+}
    =
    \frac{1}{\sqrt{2}}
    \left(
    D_i s^i + K_{ij}s^i s^j - K
    \right),
\end{equation}
where \(D_i\) is the covariant derivative compatible with \(\gamma_{ij}\).

A \textit{marginally outer trapped surface} (MOTS) is a smooth closed two-surface \(S\subset\Sigma_t\) satisfying
\begin{eqnarray}
    \Theta_{+}=0.
\end{eqnarray}
The MOTS equation is a nonlinear elliptic equation for the shape of the surface.  This elliptic property is central both to apparent-horizon finding in numerical relativity and to the stability theory \cite{Thornburg:2007eventAHfinders,Andersson:2007fh}. Below we write \(\Theta\equiv\Theta_+\) when no confusion can arise.
The \textit{apparent horizon} on a slice \(\Sigma_t\) is the outermost MOTS on that slice, equivalently the outer boundary of the trapped region when such a boundary exists.  

A \textit{quasilocal horizon} or \textit{marginally outer trapped tube} (MOTT) is a three-dimensional hypersurface in spacetime foliated by MOTSs. We label the MOTS leaves of a continuous MOTT by a smooth parameter \(\tau\),
\begin{eqnarray}
    \mathcal H=\bigcup_{\tau\in I}S_\tau,
\end{eqnarray}
where \(\tau\) need not coincide with the Cauchy time \(t\).  Relative to the chosen foliation, the same continuous MOTT may be tangent to \(\Sigma_{t_*}\) at \(S_*\).  For sufficiently small \(t>t_*\), it can then intersect a single slice \(\Sigma_t\) in two MOTS leaves, \(S_+(t)\) and \(S_-(t)\), which are the outer and inner common horizon branches.  The causal character of \(\mathcal{H}\) is not fixed in general. A MOTT may be spacelike, null, timelike, or of mixed signature \cite{BoothEtAl:2006MTT}. For the smooth outer branch, strict stability and the null energy condition restrict the tube to be locally spacelike or null \cite{Andersson:2007fh}.

The cross-sections \(S_\tau\) of a quasilocal horizon carry intrinsic and extrinsic geometric data, such as the area
\begin{eqnarray}
    A[S_\tau] = \int_{S_\tau} dA.
\end{eqnarray}
These quantities are especially useful in numerical relativity because they can be computed directly on the horizon during the simulation.  They also provide a way to compare strong-field dynamics near the black hole with radiation measured in the wave zone or at null infinity.

\begin{figure}[tbp]
    \centering
    \includegraphics[width=0.95\columnwidth]{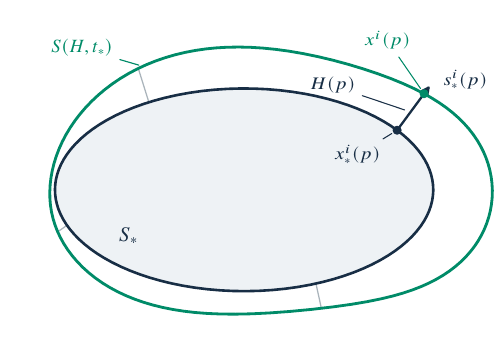}
    \caption{Normal-graph representation of a nearby surface on the reference slice $\Sigma_{t_*}$. The graph $S(H,t_*)$ is obtained by displacing each point $x_*^i(p)$ of $S_*$ by $H(p)s_*^i(p)$ in the chosen coordinates, along the fixed reference normal $s_*^i$.}
    \label{fig:reference-normal-graph}
\end{figure}
\begin{figure}[t]
    \centering
    \includegraphics[width=0.97\columnwidth]{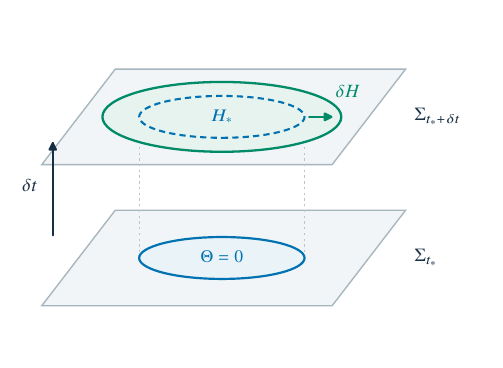}
    \caption{Changes in the time slice and surface shape. The reference MOTS $S_*$ lies on $\Sigma_{t_*}$. Holding $H=H_*$ fixed identifies the same coordinate surface on the later slice (dashed blue), although the slice geometry and the surface's expansion generally change. A normal-graph perturbation $\delta H$ gives a nearby MOTS (green) when one exists.}
    \label{fig:slice-shape-variation}
\end{figure}

\subsection{Representation of MOTS with a reference surface}\label{representation}
Let $S_*$ be a two-sphere in the slice $\Sigma_{t_*}$ that can be represented by a smooth level-set function $u(x^i)$ such that
\begin{equation}
    S_*=\{x^i\in\Sigma_{t_*}:u(x^i)=0\}.
\end{equation}
We use a fixed choice of spatial coordinates to identify points on nearby slices.
The sign of $u$ is chosen so that the corresponding normal points
outward. The outward unit normal to $S_*$ can be computed using the associated spatial metric $\gamma_{ij}(t_*)$, as
\begin{equation}
    s_*^i
    =
    \left.
    \frac{\gamma^{ij}(t_*)\partial_j u}
    {\sqrt{\gamma^{ij}(t_*)\partial_i u\,\partial_j u}}
    \right|_{S_*}.
    \label{eq:star-normal}
\end{equation}

Let $p\in S_*$ be a point on $S_*$, and let $x_*^i(p)$
be its coordinate position. A nearby surface in the slice $\Sigma_t$ can be described by a
scalar field $H(p)$ on $S_*$, its normal graph height, as
\begin{equation}
    S(H,t)
    =
    \left\{
    (t,x^i):
    x^i=x_*^i(p)+H(p)s_*^i(p),
    \quad
    p\in S_*
    \right\}.
\end{equation}
The scalar $H(p)$ specifies the coordinate displacement from $x_*^i(p)$ along the fixed reference direction $s_*^i(p)$ (Fig.~\ref{fig:reference-normal-graph}).

Equivalently, if $x^A=(\theta,\phi)$ is a coordinate chart on $S_*$, then
$H$ may be written as $H(x^A)$. With this convention, $S_*$ corresponds to
\begin{equation}
    H_*\equiv0,
    \qquad
    S_*=S(H_*,t_*)=S(0,t_*).
\end{equation}
We retain the symbol \(H_*\) in later formulas to mark evaluation at the reference surface.

This construction provides a way to describe surfaces on nearby slices. 
Holding $H(x^A)$ fixed while changing $t$ means evaluating the same coordinate surface on a different time slice (Fig.~\ref{fig:slice-shape-variation}), while the geometric data
$\gamma_{ij}(t)$ and $K_{ij}(t)$ change from slice to slice.

For a surface $S(H,t)$, its outgoing null expansion is given by
\begin{align}
    \Theta[H,t](p)
    ={}&
    \frac{1}{\sqrt{2}}
    \Bigl(
    D_i s^i[H,t]
    +K_{ij}(t)s^i[H,t]s^j[H,t]
    \nonumber\\
    &\qquad
    -K(t)
    \Bigr)\Bigr|_{x^i=x_*^i(p)+H(p)s_*^i(p)} .
\end{align}
The reference normal $s_*^i$ fixes the graph parametrization; the current normal $s^i[H,t]$ is recomputed on every trial surface using the metric on its slice. The graph height is a function on the fixed reference sphere,
\begin{equation}
    H \in X, \quad H : S^2 \rightarrow \mathbb{R},
\end{equation}
and the outgoing expansion defines the map
\begin{eqnarray}
    \Theta[\;\cdot\;,\;\cdot\;] : X\, \times\, \mathbb{R} \rightarrow Y,
\end{eqnarray}
which assigns to each pair $(H,t)$ a function on the same sphere,
\begin{equation}
    \Theta[H,t] \in Y, \quad \Theta[H,t] : S^2 \rightarrow \mathbb{R}.
\end{equation}
Thus $\Theta[H,t]$ is the pointwise outgoing null expansion evaluated on the surface labeled by $H$. 

We now define the functional derivatives with respect to the surface shape. For treatments of surface deformations, see Refs.~\cite{BoothFairhurst:2007deformations,BoothKunduriOGrady:2017unstableMOTS}. For any sufficiently smooth quantity $F[H,t]$ with the appropriate target space, perturbing $H$ in the direction $V\in X$ gives the first functional derivative
\begin{equation}
    \left.\delta_HF\right|_{(H,t)}[V]
    :=
    \left.
    \frac{d}{d\epsilon}
    F[H+\epsilon V,t]
    \right|_{\epsilon=0}.
\end{equation}
Perturbing $H$ independently in several directions gives the second and higher functional derivatives:
\begin{widetext}
\begin{align}
    \left.\delta_H\delta_HF\right|_{(H,t)}[V_1,V_2]
    :={}&
    \left.
    \frac{\partial^2}{\partial\epsilon_1\partial\epsilon_2}
    F[H+\epsilon_1V_1+\epsilon_2V_2,t]
    \right|_{\epsilon_1=\epsilon_2=0},
    \\
    \left.
    \underbrace{\delta_H\cdots\delta_H}_{n\ {\rm times}}F
    \right|_{(H,t)}
    [V_1,\ldots,V_n]
    :={}&
    \left.
    \frac{\partial^n}{\partial\epsilon_1\cdots\partial\epsilon_n}
    F\!\left[H+\sum_{a=1}^n\epsilon_aV_a,t\right]
    \right|_{\epsilon_1=\cdots=\epsilon_n=0}.
\end{align}
\end{widetext}
Here $V_1,\ldots,V_n\in X$. The square brackets list the perturbations on which the functional derivative acts, while $t$ is held fixed. For sufficiently smooth $F$, the higher functional derivatives are symmetric in these perturbations. 

On a segment where the Cauchy time \(t(\tau)\) varies monotonically along the continuous MOTT, its leaves can be represented by a single graph \(H(t)\) satisfying
\begin{equation}
    \Theta[H(t),t]=0.
\end{equation}
Near the turning point \(t=t_*\), the same smooth MOTT is represented on later slices by two graph branches \(H_+(t)\) and \(H_-(t)\), satisfying
\begin{equation}
    \Theta[H_\pm(t),t]=0,
    \qquad H_\pm(t_*)=H_*.
\end{equation}
Thus the labels \(+\) and \(-\) distinguish two intersections with the same continuous MOTT.

For readability, all derivatives of $\Theta$ appearing below are
evaluated at the reference point $(H_*,t_*)$, unless a different point is
explicitly stated. We use the shorthand
\begin{equation}
    \Theta_*:=\Theta[H_*,t_*],
\end{equation}
\begin{equation}
    L_\Sigma[\,\cdot\,]
    :=
    \left.
    \delta_H\Theta
    \right|_{(H_*,t_*)}
    [\,\cdot\,],
\end{equation}
\begin{equation}
    B_\Sigma[\,\cdot\,,\,\cdot\,]
    :=
    \left.
    \delta_H\delta_H\Theta
    \right|_{(H_*,t_*)}
    [\,\cdot\,,\,\cdot\,],
\end{equation}
\begin{equation}
    \Theta_t
    :=
    \left.
    \partial_t\Theta
    \right|_{(H_*,t_*)},
    \qquad
    \Theta_{tt}
    :=
    \left.
    \partial_t^2\Theta
    \right|_{(H_*,t_*)},
\end{equation}
and
\begin{equation}
    \dot L_\Sigma[\,\cdot\,]
    :=
    \left.
    \partial_t(\delta_H\Theta)
    \right|_{(H_*,t_*)}
    [\,\cdot\,].
\end{equation}
Here, $L_\Sigma$ is the first-order variation of the MOTS residual with respect
to the surface shape within the three-dimensional slice. The bilinear operator $B_\Sigma$ is
the second variation with respect to the surface shape. The quantity
$\Theta_t$ is the change of the residual under a change of slice while
the coordinate shape $H$ is held fixed, and $\dot L_\Sigma$ is the
corresponding time derivative of the first-order variation operator. Square
brackets indicate the perturbation, or perturbations, on which the resulting linear or multilinear operator acts.

Let $S_*$ be chosen to be a MOTS in the slice $\Sigma_{t_*}$, so that
\begin{equation}
    \Theta[H_*,t_*]=0.
\end{equation}
We can seek nearby MOTSs by varying $H$ and $t$ around $(H_*,t_*)$,
\begin{equation}
    H=H_*+\delta H,
    \qquad
    t=t_*+\delta t.
\end{equation}
The outgoing null expansion equation can then be expanded near the
reference slice as
\begin{equation}\label{expansion1}
\begin{aligned}
    \Theta[H_*+\delta H,t_*+\delta t]
    ={}&
    \Theta_*
    +
    L_\Sigma[\delta H]
    +
    \delta t\,\Theta_t
    +
    \frac{1}{2}B_\Sigma[\delta H,\delta H]
    \\
    &+
    \delta t\,\dot L_\Sigma[\delta H]
    +
    \frac{1}{2}(\delta t)^2\Theta_{tt}
    + \cdots.
\end{aligned}
\end{equation}

\subsection{The MOTS stability operator}
The operator
\begin{equation}
    L_\Sigma[\,\cdot\,] : X \rightarrow Y
\end{equation}
is referred to as the MOTS stability operator. Its spectral properties have been studied extensively in the MOTS literature, in particular in the work of Andersson, Mars, and Simon \cite{Andersson:2007fh}.
It has a real principal eigenvalue \(\lambda_0\), whose eigenfunction can be chosen to be strictly positive. Stability of a MOTS is characterized by the sign of this principal eigenvalue; stability corresponds to \(\lambda_0\geq 0\), while strict stability corresponds to \(\lambda_0>0\). 

A useful result is that the apparent horizon, the outermost MOTS, on a slice must be stable, i.e., \(\lambda_0\geq 0\). Otherwise, a first-order outward deformation would produce an outer trapped surface outside it, contradicting the assumption that the surface is outermost.

\section{Local structure of common horizon formation}\label{sec:theory}
In this section, we derive the local structure of common horizon formation through linear order in time, including both the branch separation and the shared linear drift that give the form of a tilted parabola. We first establish that the remnant horizon
appears as a new common MOTS enclosing the two progenitor horizons, rather than
forming by continuous deformation of the individual horizons. We set
up a perturbative description of the MOTS equation about the formation
leaf. We then show that past isolation forces the MOTS stability operator $L_\Sigma$ to have a zero mode, and that outermost stability makes its kernel one-dimensional. We utilize this property through the Lyapunov--Schmidt reduction that reduces the infinite-dimensional MOTS equation to a bifurcation normal form, from which the two horizon branches and a square-root scaling law for quasilocal horizon quantities follow.
\subsection{Formation of the common horizon}
\begin{figure*}[t]
    \centering
    \includegraphics[width=0.98\textwidth]{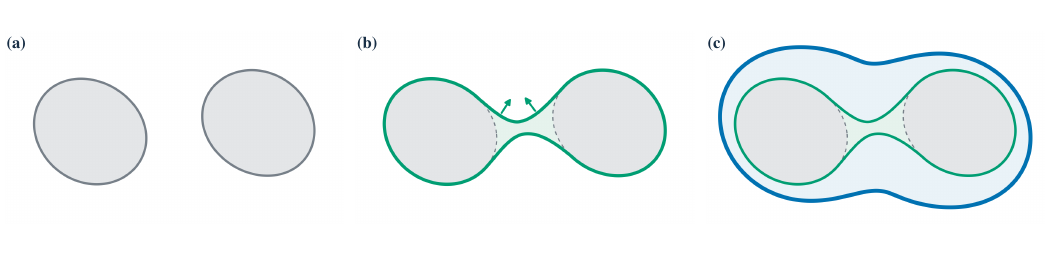}
    \caption{The neck-surgery construction behind the jump of the outermost MOTS. (a) Two individual MOTSs (gray). (b) The surgery surface (green) coincides with the individual MOTSs away from the gap and joins them by a narrow, approximately hyperboloidal neck. Dashed gray arcs indicate the portions removed. At the neck waist, the outward normals $s^i$ converge and $D_i s^i\sim-1/\rho$, where $\rho$ is the waist curvature radius. For sufficiently small separation, the neck is weakly outer trapped. (c) An enclosing outermost MOTS (blue) bounds the resulting connected trapped region (shaded), so the apparent horizon jumps outward before the individual horizons touch \cite{AnderssonMetzger:2009,Andersson:2008up}.}
    \label{fig:neck-surgery}
\end{figure*}
Under the hypotheses of Ref.~\cite{Andersson:2008up}, when two components of an outermost MOTS approach sufficiently closely, the outermost MOTS must jump before the components touch, and a connected trapped region containing both components is produced.  The following curvature estimate gives the geometric intuition behind that construction, illustrated in Fig.~\ref{fig:neck-surgery}.  In the narrow neck joining the two MOTSs, the inserted surface is approximately hyperboloidal and has negative mean curvature whose leading magnitude scales inversely with the neck curvature radius when the waist is kept wider than that radius,
\begin{equation}
    D_i s^i \sim -\frac{1}{\rho}.
\end{equation}
The curvature radius \(\rho\) scales with the separation between the MOTSs and tends to zero as contact is approached; equivalently, the curvature \(1/\rho\) blows up.

Since the outgoing null expansion of a surface is given by
\begin{equation}
    \Theta_{+}=\frac{1}{\sqrt{2}}\left(D_i s^i +K_{ij}s^i s^j - K\right),
\end{equation}
and the last two terms are associated with the background curvature and remain finite, sufficiently decreasing the separation between the two MOTSs causes the negative mean-curvature term to dominate. In this regime,
\begin{equation}
    |D_i s^i| \geq \left|K_{ij}s^i s^j - K\right|
\end{equation}
everywhere on the neck.  Since \(D_i s^i<0\) there, this bound implies
\begin{equation}
    \Theta_+=\frac{1}{\sqrt{2}}\left(D_i s^i+K_{ij}s^i s^j-K\right)\leq0.
\end{equation}
The neck is therefore weakly outer trapped.  This estimate is only the local geometric intuition; the rigorous surgery construction controls and smooths the complete connected surface, while the trapped-region existence result supplies the enclosing outermost MOTS \cite{AnderssonMetzger:2009,Andersson:2008up}.

\subsection{Properties of the linear response}
We choose the reference surface $S_*$ to be the common MOTS at first formation in the chosen foliation. More precisely, $S_*$ is taken to be past isolated: in a sufficiently small spacetime neighborhood
of $S_*$, there is no MOTS of the common-horizon branch on leaves
immediately before $t_*$. 

Suppose, for contradiction, that $L_\Sigma$ is invertible. Consider a
nearby slice
\begin{equation}
    t=t_*+\delta t,
    \qquad
    |\delta t|\ll 1 .
\end{equation}
Keeping the reference shape fixed, and since $\Theta_*=0$, the nearby expansion in
Eq.~\eqref{expansion1} gives
\begin{equation}
    \Theta[H_*,t_*+\delta t]
    =
    \delta t\,\Theta_t
    +
    O((\delta t)^2).
\end{equation}
If $L_\Sigma$ is invertible, the leading residual can be cancelled by a
first surface correction
\begin{equation}
    \delta H^{(1)}
    =
    -
    L_\Sigma^{-1}
    \left[
        \delta t\,\Theta_t
    \right],
\end{equation}
\begin{equation}
    L_\Sigma[\delta H^{(1)}]
    +
    \delta t\,\Theta_t
    =
    0 .
\end{equation}

Since $\delta H^{(1)}=O(\delta t)$, substituting
$H=H_*+\delta H^{(1)}$ back into the expansion cancels the MOTS residual to
first order, leaving only terms of order $(\delta t)^2$. Smoothness of $\Theta$ and invertibility of $L_\Sigma$ ensure that it extends uniquely to an exact nearby solution,
\begin{equation}
    \Theta[H(t_*+\delta t),t_*+\delta t]=0 .
\end{equation}

This construction does not distinguish between positive and negative
$\delta t$. Thus, an invertible $L_\Sigma$ would imply a local MOTT
passing through $S_*$ and extending to both sides of the reference slice.
This contradicts the past isolation of the common horizon. Therefore, $L_\Sigma$ cannot be invertible at the
formation leaf. 

For this elliptic operator on the closed surface $S_*$, failure of invertibility implies the existence of a nontrivial zero mode.
\begin{equation}
    \ker L_\Sigma\neq\{0\}.
\end{equation}

We now identify the kernel. The MOTS stability
operator is a second-order elliptic operator on the closed surface $S_*$.
It is not self-adjoint in general, so its spectrum need not be real.
However, it has a real principal eigenvalue $\lambda_0$ satisfying
\begin{equation}
    \operatorname{Re}\lambda_n\geq \lambda_0
\end{equation}
for every eigenvalue $\lambda_n$. The corresponding principal
eigenfunction can be chosen strictly positive and is unique up to an
overall normalization \cite{Andersson:2007fh}.

Since $L_\Sigma$ has a kernel, zero is an eigenvalue. Therefore
\begin{equation}
    \lambda_0\leq 0 .
\end{equation}
On the other hand, $S_*$ is taken to be the outermost MOTS on the formation slice. Outermost MOTSs are stable, and this stability is equivalent to non-negativity of the
principal eigenvalue \cite{Andersson:2007fh}. Hence
\begin{equation}
    \lambda_0\geq 0 .
\end{equation}
Combining these two inequalities gives
\begin{equation}
    \lambda_0=0 .
\end{equation}

Thus, the zero eigenvalue is the principal eigenvalue of $L_\Sigma$. We
denote the corresponding eigenfunction by $H_0(x^A)$,
\begin{equation}
    L_\Sigma[H_0]=0,
    \qquad
    H_0>0 .
\end{equation}

It remains to exclude additional independent zero modes. Let $G_0\in\ker L_\Sigma$.
Since $H_0$ is strictly positive on the connected surface $S_*$, we may
write
\begin{equation}
    G_0=f H_0
\end{equation}
for some smooth function $f$ on $S_*$. Substituting this into
$L_\Sigma[G_0]=0$ and using $L_\Sigma[H_0]=0$, the equation reduces to an elliptic equation involving only derivatives of $f$ on the closed surface $S_*$. The maximum-principle result of Andersson, Mars, and Simon then implies that
$f$ must be constant \cite{Andersson:2007fh}. Hence, the kernel  is one-dimensional,
\begin{equation}
    \ker L_\Sigma=\operatorname{span}\{H_0\}.
\end{equation}

\subsection{Local branch structure}

Since the kernel of $L_\Sigma$ is one-dimensional, the surface
perturbation can be decomposed into a single kernel direction and a
complementary part,
\begin{equation}
    \delta H=\mathcal{A} H_0+\delta H_\perp .
\end{equation}
Here $\mathcal{A}$ is the amplitude along the kernel direction, while
$\delta H_\perp$ is the complementary surface deformation. Since
\begin{equation}
    L_\Sigma[H_0]=0,
\end{equation}
we have
\begin{equation}
    L_\Sigma[\delta H]=L_\Sigma[\delta H_\perp].
\end{equation}

Let $dA_*$ be the surface element induced on $S_*$ by $\gamma_{ij}(t_*)$, and define the $L^2$ pairing on $S_*$ by
\begin{equation}
    \langle f|g\rangle
    :=
    \int_{S_*}fg\,dA_*.
\end{equation}
We choose the normalization
\begin{equation}
    \langle H_0|H_0\rangle=1,
\end{equation}
and impose
\begin{equation}
    \langle H_0|\delta H_\perp\rangle=0 .
\end{equation}
Thus
\begin{equation}
    \mathcal{A}=\langle H_0|\delta H\rangle,
    \qquad
    \delta H_\perp=\delta H-\mathcal{A} H_0 .
\end{equation}
This split of the perturbation also induces a natural splitting of the space of graph functions; defining
\begin{equation}
    X_\perp
    :=
    \{f\in X:\langle H_0|f\rangle=0\},
\end{equation}
we have
\begin{equation}
    X=\operatorname{span}\{H_0\}\oplus X_\perp,
    \qquad
    \delta H_\perp\in X_\perp .
\end{equation}

For readability and order consistency, we adopt an \textit{a priori} scaling for $\delta t$ and $\mathcal{A}$ near the reference leaf; this order counting will be justified at the end of the reduction. We take
\begin{equation}
    \mathcal{A}=O(\epsilon),
    \qquad
    \delta t=O(\epsilon^2).
\end{equation}
We retain terms through $O(\epsilon^3)$ and verify the assumed scaling after solving the projected equations. The third shape variation entering at that order is denoted
\begin{equation}
    T_\Sigma[\,\cdot\,,\,\cdot\,,\,\cdot\,]
    :=
    \left.
    \delta_H\delta_H\delta_H\Theta
    \right|_{(H_*,t_*)}
    [\,\cdot\,,\,\cdot\,,\,\cdot\,].
\end{equation}

Under this splitting and counting, the expansion~\eqref{expansion1}
organizes by order in $\epsilon$,
\begin{widetext}
    \begin{equation}\label{eq:theta-expanded-split}
\begin{aligned}
    \Theta[H_*+\mathcal{A} H_0+\delta H_\perp,t_*+\delta t]
    ={}&
    \underbrace{
        L_\Sigma[\delta H_\perp]
        +\delta t\,\Theta_t
        +\frac{1}{2}\mathcal{A}^2 B_\Sigma[H_0,H_0]
    }_{O(\epsilon^2)}
    \\
    &+
    \underbrace{
        \mathcal{A}\delta t\,\dot L_\Sigma[H_0]
        +\mathcal{A} B_\Sigma[H_0,\delta H_\perp]
        +\frac{1}{6}\mathcal{A}^3 T_\Sigma[H_0,H_0,H_0]
    }_{O(\epsilon^3)}
    +O(\epsilon^4).
\end{aligned}
\end{equation}
\end{widetext}

Here we used $\Theta_*=0$ and $L_\Sigma[H_0]=0$.

To find the nearby MOTS, we must solve $\Theta = 0$ in
Eq.~\eqref{eq:theta-expanded-split} for $\delta H_\perp$ and the kernel
amplitude $\mathcal{A}$. The unknown $\delta H_\perp$ enters through
$L_\Sigma[\delta H_\perp]$, but $L_\Sigma$ has a kernel spanned by $H_0$ and so
cannot be inverted directly. We will split the domain and target spaces below so that the restriction of $L_\Sigma$ to $X_\perp$ is invertible and determines $\delta H_\perp$. Substituting this result into the remaining equation, the projection onto the cokernel direction, gives a scalar equation for $\mathcal{A}$. This procedure is called Lyapunov--Schmidt reduction \cite{Potzsche2011BifurcationTheory,Kielhofer:2012bt}, a standard tool in bifurcation theory.

By Lemma~4.1 of Andersson, Mars, and Simon
\cite{Andersson:2007fh}, the adjoint operator
$L_\Sigma^\dagger$ has the same principal eigenvalue as
$L_\Sigma$ and, if $\lambda_0=0$, the adjoint has a
positive principal zero mode $H_0^\dagger$, unique up to
normalization. We choose
\begin{equation}
    L_\Sigma^\dagger[H_0^\dagger]=0,
    \qquad
    \langle H_0^\dagger|H_0^\dagger\rangle=1,
\end{equation}
where $L_\Sigma^\dagger[\;\cdot\;]$ is the $L^2$ adjoint of
$L_\Sigma[\;\cdot\;]$.

For any $f\in X$, the definition of the adjoint gives
\begin{equation*}
\begin{aligned}
    \langle H_0^\dagger|L_\Sigma[f]\rangle
    &=
    \langle L_\Sigma^\dagger[H_0^\dagger]|f\rangle
    \\
    &=0.
\end{aligned}
\end{equation*}
Thus every element in the range of $L_\Sigma$ is orthogonal
to $H_0^\dagger$,
\begin{equation*}
    \text{Range}(L_\Sigma)
    \subseteq
    \left\{
        g\in Y:
        \langle H_0^\dagger|g\rangle=0
    \right\}.
\end{equation*}
Conversely, Lemma~4.3 of Ref.~\cite{Andersson:2007fh},
specialized to the principal eigenvalue $\lambda_0=0$, implies
that every $g\in Y$ satisfying
$\langle H_0^\dagger|g\rangle=0$ can be written as
$g=L_\Sigma[f]$ for some $f$.
Therefore,
\begin{equation*}
    \text{Range}(L_\Sigma)
    =
    \left\{
        g\in Y:
        \langle H_0^\dagger|g\rangle=0
    \right\}.
\end{equation*}

Every $g\in Y$ can be
decomposed uniquely into its component along $H_0^\dagger$
and a component in $\text{Range}(L_\Sigma)$. Therefore, the
function space $Y$ admits the splitting
\begin{equation}
    Y
    =
    \operatorname{span}\{H_0^\dagger\}
    \oplus
    \text{Range}(L_\Sigma).
\end{equation}

We now solve for $\delta H_\perp$ by inverting $L_\Sigma$ on its range.
Define the cokernel projection
\begin{eqnarray}
    \mathbb{P} : Y \rightarrow \operatorname{span}\{H_0^\dagger\}
\end{eqnarray}
and complementary projection
\begin{eqnarray}
    \mathbb{Q} \equiv \mathbb{I} - \mathbb{P} : Y \rightarrow \text{Range}(L_\Sigma),
\end{eqnarray}
with explicit form
\begin{equation}
    \mathbb{P}u:=\langle H_0^\dagger|u\rangle H_0^\dagger,
\end{equation}
and 
\begin{equation}
   \mathbb{Q}u:=u-\langle H_0^\dagger|u\rangle H_0^\dagger .
\end{equation}
After applying the projection, $\mathbb{P}\Theta$ and $\mathbb{Q}\Theta$ can be thought of as functions of the amplitude of the zero mode $\mathcal{A}$, the complementary deformation $\delta H_\perp$, and $t$. The projected expansion maps are 
\begin{eqnarray}
    \mathbb{P}\Theta &:& \mathbb{R} \times X_\perp \times \mathbb{R} \rightarrow \operatorname{span}\{H_0^\dagger\},\\
    \mathbb{Q}\Theta &:& \mathbb{R} \times X_\perp \times \mathbb{R} \rightarrow \text{Range}(L_\Sigma).
\end{eqnarray}
The MOTS equation is therefore equivalent to
\begin{equation}\label{eq:peq}
    \mathbb{P}\Theta[H_*+\mathcal{A} H_0+\delta H_\perp,t_*+\delta t]=0,
\end{equation}
and
\begin{equation}\label{eq:qeq}
    \mathbb{Q}\Theta[H_*+\mathcal{A} H_0+\delta H_\perp,t_*+\delta t]=0.
\end{equation}
We first determine the complementary deformation $\delta H_\perp$ from the second equation, then substitute it into the first.

\begin{figure}[tbp]
    \centering
    \includegraphics[width=0.97\columnwidth]{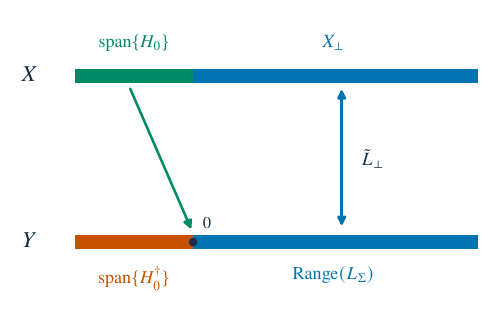}
    \caption{Decomposition of the domain $X$ and target $Y$ at formation. $L_\Sigma$ maps the kernel $\operatorname{span}\{H_0\}$ to zero, while its restriction $\tilde L_\perp$ maps $X_\perp$ invertibly onto $\operatorname{Range}(L_\Sigma)$. The adjoint zero mode $H_0^\dagger$ spans the complementary direction in $Y$.}
    \label{fig:function-space-split}
\end{figure}

After applying the $\mathbb{Q}$ projection, every source in $\text{Range}(L_\Sigma)$ has a unique solution in $X_\perp$. Define the restricted
operator
\begin{equation}
    \tilde L_\perp
    :=
    L_\Sigma|_{X_\perp}:X_\perp\rightarrow  \text{Range}(L_\Sigma) .
\end{equation}
This map is injective because $X_\perp$ meets the kernel only at zero, and surjective because the kernel component of any preimage contributes nothing to $L_\Sigma[x]$, so every element of the range is already attained from $X_\perp$ alone. Therefore, $\tilde L_\perp$ is invertible (Fig.~\ref{fig:function-space-split}), with inverse 
\begin{equation}
    \tilde L_\perp^{-1}: \text{Range}(L_\Sigma)\rightarrow X_\perp .
\end{equation}

Applying the inverse to the terms of order $\epsilon^2$ in Eq.~\eqref{eq:qeq} gives
\begin{equation}
    \delta H_\perp(\mathcal{A},\delta t)
    =
    \delta t\,H_t
    +
    \mathcal{A}^2H_2
    +
    O(\epsilon^3),
\end{equation}
where
\begin{equation}
    H_t
    :=
    -\tilde L_\perp^{-1}\mathbb{Q}[\Theta_t],
\end{equation}
and
\begin{equation}
    H_2
    :=
    -\frac{1}{2}\tilde L_\perp^{-1}\mathbb{Q}[B_\Sigma[H_0,H_0]] .
\end{equation}
Therefore
\begin{equation}
    \delta H_\perp
    =
    O(\delta t+\mathcal{A}^2)
    =
    O(\epsilon^2),
\end{equation}
which justifies the order counting assumed above.

Substituting $\delta H_\perp(\mathcal{A},\delta t)$ into Eq.~\eqref{eq:peq},
\begin{equation}
    \left\langle
    H_0^\dagger
    \middle|
    \Theta[H_*+\mathcal{A} H_0+\delta H_\perp(\mathcal{A},\delta t),
    t_*+\delta t]
    \right\rangle
    =0 ,
\end{equation}
and using
$\langle H_0^\dagger|L_\Sigma[\delta H_\perp]\rangle
=\langle L_\Sigma^\dagger[H_0^\dagger]|\delta H_\perp\rangle=0$, the
linear term drops. The reduced scalar equation through $O(\epsilon^3)$ is therefore
\begin{equation}\label{eq:tilted-normal-form}
    0
    =
    c_t\,\delta t
    +
    c_2\,\mathcal{A}^2
    +
    c_{1t}\,\mathcal{A}\delta t
    +
    c_3\,\mathcal{A}^3
    +
    O(\epsilon^4),
\end{equation}
with
\begin{equation}
    c_t:=\langle H_0^\dagger|\Theta_t\rangle,
    \qquad
    c_2:=\frac{1}{2}\langle H_0^\dagger|B_\Sigma[H_0,H_0]\rangle ,
\end{equation}
\begin{equation}
    c_{1t}
    :=
    \langle H_0^\dagger|\dot L_\Sigma[H_0]\rangle
    +
    \langle H_0^\dagger|B_\Sigma[H_0,H_t]\rangle ,
\end{equation}
and
\begin{equation}
    c_3
    :=
    \frac{1}{6}
    \langle H_0^\dagger|T_\Sigma[H_0,H_0,H_0]\rangle
    +
    \langle H_0^\dagger|B_\Sigma[H_0,H_2]\rangle .
\end{equation}

The derivation rests on two requirements,
\begin{equation}
    c_t\neq0,
    \qquad
    c_2\neq0,
\end{equation}
and both have direct geometric interpretations.  The first says that advancing the slice changes the expansion along the one direction the linearized operator cannot supply, so time genuinely drives the bifurcation.  The second says that the leading response of the expansion along $H_0$ is quadratic, i.e., the tangency of slice to MOTT is quadratic.

Section~\ref{sec:numerical} tests the branch behavior implied by these requirements. For $c_t\neq0$, a vanishing $c_2$ with $c_3\neq0$ would instead give the single real branch $\mathcal A\propto(\delta t)^{1/3}$. The simulations resolve two branches, and their freely fitted exponents are close to $1/2$ (Table~\ref{tab:formation-times}), supporting the quadratic leading balance.  Given $c_2\neq0$, the nonzero measured splitting amplitudes then require $c_t\neq0$.

Because the common branches exist immediately after formation, we now take $\delta t>0$. Past isolation then requires
\begin{equation}
    -\frac{c_t}{c_2}>0 .
\end{equation}
Define
\begin{equation}
    \mathcal{B}:=\sqrt{-\frac{c_t}{c_2}} .
\end{equation}
Solving the reduced equation gives
\begin{equation}
    \mathcal{A}_\pm(\delta t)
    =
    \pm \mathcal{B}(\delta t)^{1/2}
    +
    \mathcal{C}\,\delta t
    +
    O((\delta t)^{3/2}),
\end{equation}
where
\begin{equation}
    \mathcal{C}
    =
    -\frac{1}{2c_2}
    \left(
        c_{1t}+c_3\mathcal{B}^2
    \right).
\end{equation}
This verifies the assumed order counting. 

Thus, the corresponding surface perturbations are
\begin{eqnarray}\label{eq:Hexpansion}
    \delta H_\pm(\delta t)\nonumber
    &=&
    \mathcal{A}_\pm H_0
    +
    \delta H_\perp(\mathcal{A}_\pm,\delta t)\\
    &=&
    \pm \mathcal{B}(\delta t)^{1/2}H_0
    +
    \delta t H_d
    +
    O((\delta t)^{3/2}),
\end{eqnarray}
where $H_d = \mathcal{C} H_0+H_t+\mathcal{B}^2H_2$.
This is the local form of a tilted parabola. The leading $(\delta t)^{1/2}$ term gives the branch separation, while $\delta t\,H_d$ gives the shared linear drift. Together they specify the surface displacement through linear order in time. For sufficiently small $\delta t>0$, $H_0>0$ implies that the $+$ branch lies outside the reference surface and is the outer common horizon, while the $-$ branch is the inner common horizon.

The square-root exponent is unchanged by a smooth reparametrization of time whose derivative is positive at \(t_*\).  The coefficients depend on the time coordinate, slicing, and identification of nearby surfaces.  Under a rescaling $H_0\mapsto aH_0$, the coefficients transform as $\mathcal B\mapsto\mathcal B/a$ and $\mathcal C\mapsto\mathcal C/a$, leaving the displacement fields $\mathcal BH_0$, $\mathcal CH_0$, and $H_d$ unchanged. The MOTT itself also depends on the foliation. A null MOTT is the same for every foliation, because each of its cross sections is marginally trapped \cite{Ashtekar:2004cn}. A non-null MOTT is not: where it is spacelike, its foliation by marginally trapped surfaces is unique \cite{AshtekarGalloway:2005uniqueness}, so a different slicing of the same tube does not yield MOTSs, and a different foliation of spacetime in general generates a different tube. The analysis of this section does not single out a foliation. It applies to any foliation with a past-isolated, outermost MOTS at which $c_t\neq0$ and $c_2\neq0$, and the branch structure of Eq.~\eqref{eq:Hexpansion} then follows for the tube of that foliation.

The branch expansion Eq.~\eqref{eq:Hexpansion} leads to a scaling law for a generic sufficiently regular scalar field on the horizon two-sphere, provided its first functional variation along \(H_0\) does not vanish. Let $\Phi[H,t]$ be a sufficiently regular scalar field evaluated on the surface $S(H,t)$ and written in the same coordinates $x^A$ on $S_*$. Thus $\Phi[\cdot,\cdot]:X\times\mathbb{R}\rightarrow Y$, in the same sense as $\Theta[\cdot,\cdot]$. Define
\begin{equation}
    \Phi_\pm(\delta t):=\Phi[H_*+\delta H_\pm(\delta t),t_*+\delta t],\qquad \Phi_*:=\Phi[H_*,t_*].
\end{equation}
Expanding $\Phi[H,t]$ about $(H_*,t_*)$ gives
\begin{eqnarray}
     \Phi_\pm(\delta t)&=&\Phi_*+\left.\delta_H\Phi\right|_*[\delta H_\pm]+\delta t\left.\partial_t\Phi\right|_*\nonumber\\
     &&+\,\frac12\left.\delta_H\delta_H\Phi\right|_*[\delta H_\pm,\delta H_\pm]+O((\delta t)^{3/2}).
\end{eqnarray}
   
Here terms such as $\delta t\,\partial_t\delta_H\Phi[\delta H_\pm]$ are $O((\delta t)^{3/2})$ and are included in the remainder. Substituting Eq.~\eqref{eq:Hexpansion} gives
\begin{widetext}
    \begin{align}
    \Phi_\pm(\delta t)=\Phi_*&\pm\mathcal{B}\left.\delta_H\Phi\right|_*[H_0](\delta t)^{1/2}\nonumber\\
    &+\left(\left.\partial_t\Phi\right|_*+\left.\delta_H\Phi\right|_*[H_d]+\frac12\mathcal{B}^2\left.\delta_H\delta_H\Phi\right|_*[H_0,H_0]\right)\delta t+O((\delta t)^{3/2}).
\end{align}
\end{widetext}
Thus, when \(\left.\delta_H\Phi\right|_*[H_0]\) is not identically zero, the scalar field has a square-root branch splitting together with a common linear drift.  If this first variation vanishes, the first nonvanishing higher-order variation determines the leading branch dependence.

Now let $\mathcal Q[H,t]$ be a quasilocal scalar functional,
\begin{equation}
    \mathcal Q[H,t]=\int_{S_*}\Phi[H,t]\,dA[H,t],
\end{equation}
where $dA[H,t]$ is the induced area element of $S(H,t)$ written on the reference sphere.  At the reference surface,
\begin{equation}
    \mathcal Q_*
    :=
    \mathcal Q[H_*,t_*]
    =
    \int_{S_*}\Phi_*\,dA_*.
\end{equation}
Define
\begin{equation}
    dA_\pm(\delta t)
    :=dA[H_*+\delta H_\pm(\delta t),t_*+\delta t].
\end{equation}
To expand the area element, we use the standard first variation of area,
\begin{equation}
    \left.\delta_HdA\right|_*[X]
    =(D_i s_*^i)X\,dA_*.
\end{equation}
Combining this with Eq.~\eqref{eq:Hexpansion} gives
\begin{align}
    dA_\pm(\delta t)
    ={}&dA_*
    \pm \mathcal B(D_i s_*^i)H_0dA_*(\delta t)^{1/2}
    \nonumber\\
    &+\Bigl[
        \left.\partial_tdA\right|_*
        +(D_i s_*^i)H_d\,dA_*
    \Bigr]\delta t
    \nonumber\\
    &+\frac12\mathcal B^2
    \left.\delta_H\delta_HdA\right|_*[H_0,H_0]\delta t
    +O((\delta t)^{3/2}).
\end{align}
Multiplying this expression by the expansion of $\Phi_\pm(\delta t)$ and integrating over the fixed surface $S_*$ gives
\begin{equation}\label{eq:Q-order-by-order}
    \mathcal Q_\pm(\delta t)-\mathcal Q_*
    =\pm\mathcal A_{\mathcal Q}(\delta t)^{1/2}
    +\mathcal B_{\mathcal Q}\,\delta t
    +O((\delta t)^{3/2}),
\end{equation}
where the square-root coefficient is
\begin{align}
    \mathcal A_{\mathcal Q}
    =\mathcal B\int_{S_*}
    \Bigl[
        \left.\delta_H\Phi\right|_*[H_0]
        +(D_i s_*^i)H_0\Phi_*
    \Bigr]dA_* .
\end{align}
The coefficient of $\delta t$ is
\begin{align}
    \mathcal B_{\mathcal Q}
    ={}&\int_{S_*}\Bigl[
        \left.\partial_t\Phi\right|_*
        +\left.\delta_H\Phi\right|_*[H_d]
    \Bigr]dA_*
    \nonumber\\
    &+\frac12\mathcal B^2\int_{S_*}
        \left.\delta_H\delta_H\Phi\right|_*[H_0,H_0]dA_*
    \nonumber\\
    &+\int_{S_*}\Phi_*\left.\partial_tdA\right|_*
      +\int_{S_*}(D_i s_*^i)H_d\Phi_*dA_*
    \nonumber\\
    &+\frac12\mathcal B^2\int_{S_*}\Phi_*
        \left.\delta_H\delta_HdA\right|_*[H_0,H_0]
    \nonumber\\
    &+\mathcal B^2\int_{S_*}(D_i s_*^i)H_0
        \left.\delta_H\Phi\right|_*[H_0]
        dA_*.
\end{align}
Every contribution at $O(\delta t)$ is even under exchange of the two branches and contributes to the common linear drift. In particular, the two square-root variations change sign together, so their product contributes with the same sign on both branches, $(\pm)^2=1$. The branches therefore have opposite square-root splitting and a shared linear drift. If $\mathcal A_{\mathcal Q}$ vanishes for a special surface functional or by symmetry, the first nonvanishing term at higher order gives the leading behavior.

\subsection{Branch expansion of the principal eigenvalue}
\label{sec:eigenvalue-expansion}
The principal eigenvalue is not a scalar field carried by the horizon, so Eq.~\eqref{eq:Q-order-by-order} does not apply directly. However, perturbation of the operator gives its branch expansion directly.  Along either branch, substituting
Eq.~\eqref{eq:Hexpansion} into the shape derivative of $\Theta$ gives
\begin{align}
    \left.\delta_H\Theta\right|_{\substack{
      H_*+\delta H_\pm(\delta t),\\
      t_*+\delta t}}[X]
    ={}&L_\Sigma[X]
    \nonumber\\
    &\pm\mathcal B(\delta t)^{1/2}B_\Sigma[H_0,X]
    +O(\delta t).
    \label{eq:branch-operator-expansion}
\end{align}
The derivative in Eq.~\eqref{eq:branch-operator-expansion} follows the fixed reference direction $s_*^i$. On each branch, a graph displacement $X s_*^i$ has normal component $(s_i s_*^i)X$, where $s_i$ is the current unit normal written on $S_*$. The tangential component does not change the vanishing MOTS expansion. At formation, $s_i s_*^i=1$.
Write the graph linearization as $L_\Sigma+\delta L$, the principal eigenvalue as $\lambda_0^\pm=\delta\lambda_0$, and the normal eigenfunction as $(s_i s_*^i)(H_0+\delta H_0)$. Thus $H_0+\delta H_0$ is its graph representation. We impose $\langle H_0|\delta H_0\rangle=0$ and take both corrections to be of order $(\delta t)^{1/2}$. The eigenvalue problem therefore reads
\begin{equation}
\begin{aligned}
    &\bigl(L_\Sigma+\delta L\bigr)[H_0+\delta H_0]
    \\
    &\qquad=\delta\lambda_0\,(s_i s_*^i)(H_0+\delta H_0),
\end{aligned}
\end{equation}
Since $s_i s_*^i=1+O((\delta t)^{1/2})$, the projection factor first contributes at order $\delta t$ in this equation. Keeping terms through $O((\delta t)^{1/2})$, with $L_\Sigma[H_0]=0$, gives
\begin{equation}
    \delta L[H_0]+L_\Sigma[\delta H_0]
    =\delta\lambda_0\,H_0 .
    \label{eq:first-order-eigenproblem}
\end{equation}
Pairing with $H_0^\dagger$ removes the unknown eigenfunction correction, since
\begin{equation}
    \langle H_0^\dagger|L_\Sigma[\delta H_0]\rangle
    =\langle L_\Sigma^\dagger[H_0^\dagger]|\delta H_0\rangle
    =0,
\end{equation}
Because $L_\Sigma$ is not self-adjoint, the adjoint eigenfunction must appear on the left; hence
\begin{equation}
    \delta\lambda_0
    =\frac{\langle H_0^\dagger|\,\delta L[H_0]\rangle}{\langle H_0^\dagger|H_0\rangle}.
    \label{eq:eigenvalue-perturbation-rule}
\end{equation}
The perturbation is read off Eq.~\eqref{eq:branch-operator-expansion}, $\delta L=\pm\mathcal B(\delta t)^{1/2}B_\Sigma[H_0,\,\cdot\,]+O(\delta t)$, so
\begin{equation}
    \delta\lambda_0
    =\pm\mathcal B(\delta t)^{1/2}\,
    \frac{\langle H_0^\dagger|B_\Sigma[H_0,H_0]\rangle}{\langle H_0^\dagger|H_0\rangle}
    +O(\delta t).
\end{equation}
The pairing in the numerator is exactly twice the quadratic coefficient $c_2$ of the reduction, so the square-root coefficient of the eigenvalue is
\begin{equation}
    \mathcal A_{\lambda_0}
    =\frac{\mathcal B\,\langle H_0^\dagger|B_\Sigma[H_0,H_0]\rangle}
    {\langle H_0^\dagger|H_0\rangle}
    =\frac{2c_2\mathcal B}{\langle H_0^\dagger|H_0\rangle}.
    \label{eq:lambda-square-root-coefficient}
\end{equation}
The correction $\delta H_0$ in graph variables can be written down as well.  With $\delta\lambda_0$ known, Eq.~\eqref{eq:first-order-eigenproblem} becomes an equation for $\delta H_0$ alone, and applying the inverse $\tilde L_\perp^{-1}$ of the reduction solves it:
\begin{equation}
    \delta H_0
    =\pm(\delta t)^{1/2}
    \left(2\mathcal B\,H_2
    +\mathcal A_{\lambda_0}\,\tilde L_\perp^{-1}\mathbb Q[H_0]\right)
    +O(\delta t),
    \label{eq:eigenfunction-correction}
\end{equation}
so the correction in graph variables involves the same field $H_2$ that enters the surface deformation.

At order $\delta t$ the square-root factors enter in pairs, including the product of the leading eigenvalue with the first variation of $s_i s_*^i$. Their signs give $(\pm)^2=1$, so the linear term is shared between the branches, just as for $\mathcal B_{\mathcal Q}$ above, and the expansion closes as
\begin{equation}
    \lambda_0^\pm
    =\pm\mathcal A_{\lambda_0}(\delta t)^{1/2}
    +\mathcal B_{\lambda_0}\,\delta t
    +O\!\left((\delta t)^{3/2}\right),
    \label{eq:lambda0-slope}
\end{equation}
with $\mathcal B_{\lambda_0}$ determined by the terms of order $\delta t$; neither it nor Eq.~\eqref{eq:eigenfunction-correction} is needed in what follows.

Outermost stability also determines the signs of the reduced coefficients individually; past isolation fixes their ratio.  Since $H_0>0$, the $+$ branch lies outside and is the apparent horizon, so outermost stability (Sec.~\ref{sec:basic}) demands $\lambda_0^+\geq0$ for small $\delta t$.  Because the square-root term dominates for small $\delta t$, the sign of $\lambda_0^+$ there is the sign of $\mathcal A_{\lambda_0}$, so the bound gives $\mathcal A_{\lambda_0}\geq0$.  Both $H_0$ and $H_0^\dagger$ are positive, so $\langle H_0^\dagger|H_0\rangle>0$, and with $\mathcal B>0$, Eq.~\eqref{eq:lambda-square-root-coefficient} says $\mathcal A_{\lambda_0}$ carries the sign of $c_2$.  Since $c_2\neq0$, both are positive, and $\mathcal B^2=-c_t/c_2>0$ implies $c_t<0$.

\section{Numerical simulations}
\label{sec:numerical}

We test the consequences of the local structure derived in Sec.~\ref{sec:theory}: that on each nearby slice the two common horizons are the intersections of a single folded MOTT, with an opposite-sign displacement scaling as $(t-t_*)^{1/2}$ and a shared displacement scaling as $t-t_*$; and that quasilocal quantities inherit the same structure, a square-root splitting between the branches and a common linear drift.

\subsection{Numerical setup}

We consider three binary black hole systems: an equal-mass, nonspinning quasicircular system (Simulation~I), a nonspinning quasicircular system with mass ratio $q=2$ (Simulation~II), and an eccentric, precessing system with mass ratio $q=2$ (Simulation~III).  All dimensional quantities are normalized by $M=m_1+m_2$, the sum of the initial puncture ADM masses. The initial data and numerical parameters are given in Table~\ref{tab:nr-setup}.

The spacetime data are generated with the \texttt{Einstein Toolkit} \cite{EinsteinToolkit:2025Kruskal,Loffler:2011ay} on the \texttt{Cactus} computational infrastructure \cite{Goodale:2002a}.  Initial data are constructed with \texttt{TwoPunctures} \cite{Ansorg:2004ds}, which solves the Hamiltonian constraint for Bowen--York puncture data \cite{Bowen:1980yu,Brandt:1997tf}.  The punctures are placed on the coordinate $x$ axis with the center of mass at the origin.  For Simulations~I and II, the radial and tangential momenta are estimated with the post-Newtonian tools implemented in \texttt{NRPyPN} \cite{Healy:2017quasicircular,RamosBuades:2019eccentricity}.  Simulation~III is initialized at apastron, with its orbital parameters estimated using \texttt{SEOBNRv5EHM}\footnote{\texttt{SEOBNRv5EHM} is an eccentric, aligned-spin model, used here only to choose the initial orbital parameters.  The spins of Simulation~III lie in the orbital plane, so $\chi_{\rm eff}=0$, and the model is evaluated with vanishing spin, which is adequate for that purpose.} \cite{Gamboa:2025SEOBNRv5EHM}.  The evolution infrastructure follows the public \texttt{Einstein Toolkit} binary black hole configuration \cite{Wardell:2016gw150914}, with the initial data and grid parameters specified in Table~\ref{tab:nr-setup}.

The spacetime variables are evolved with a finite-difference implementation of the BSSN formulation \cite{Shibata:1995we,Baumgarte:1998te}.  We use the moving-puncture gauge, combining an advective $1+\log$ lapse with a Gamma-driver shift \cite{Alcubierre:2002kk,Campanelli:2005dd,Baker:2005vv}, explicit fourth-order Runge--Kutta time integration, and Kreiss--Oliger dissipation.  Mesh refinement is supplied by \texttt{Carpet} \cite{Schnetter:2003rb}.  Moving refinement centers follow the individual punctures, while another center is added at the initial center of mass to resolve the common horizon region.  The Cartesian near zone is joined to a spherical multipatch wave zone, allowing the outer boundary and waveform-extraction spheres to be placed far from the strong-field region.  Reflection symmetry is used for the nonprecessing configurations, whereas Simulation~III evolves the full domain.

\begin{table*}[t]
\footnotesize
\setlength{\tabcolsep}{2.8pt}
\renewcommand{\arraystretch}{1.18}
\caption{Initial-data and numerical parameters for Simulations~I--III.  The listed puncture momentum is $\mathbf P_1/M$ in the initial Cartesian frame, with $\mathbf P_2=-\mathbf P_1$.  Refinement entries are ordered as puncture 1, puncture 2, and common horizon.}
\label{tab:nr-setup}
\begin{ruledtabular}
\begin{tabular}{lccc}
Setting & \multicolumn{1}{c}{Simulation I} & \multicolumn{1}{c}{Simulation II} & \multicolumn{1}{c}{Simulation III} \\
\colrule
\multicolumn{4}{c}{\textit{Initial data}} \\
Mass ratio $q=m_1/m_2\geq1$ & $1$ & $2$ & $2$ \\
Initial dimensionless spins $\boldsymbol\chi_1=\boldsymbol\chi_2$ & $(0,0,0)$ & $(0,0,0)$ & $(0.4,0,0)$ \\
Initial separation $D/M$ & $11.12715149$ & $10.80383301$ & $21.94708103$ \\
Puncture momentum $\mathbf P_1/M$ & $\begin{pmatrix}-6.96664\times10^{-4}\\ 8.94926\times10^{-2}\\ 0\end{pmatrix}$ & $\begin{pmatrix}-6.12220\times10^{-4}\\ 8.12270\times10^{-2}\\ 0\end{pmatrix}$ & $\begin{pmatrix}-1.78272\times10^{-4}\\ 3.95866\times10^{-2}\\ 0\end{pmatrix}$ \\

\colrule
\multicolumn{4}{c}{\textit{Grid structure}} \\
Cells across smallest finest-grid radius $n$ & $36$ & $36$ & $32$ \\
Domain symmetry & Equatorial reflection & Equatorial reflection & Full \\
Refinement levels $(N_1,N_2,N_c)$ & $(7,7,7)$ & $(6,7,7)$ & $(6,7,7)$ \\
Finest refinement radii $(R_1,R_2,R_c)/M$ & $(0.6,0.6,1.2)$ & $(0.8,0.4,1.2)$ & $(0.8,0.4,1.2)$ \\
Finest spacings $(h_1,h_2,h_c)/M$ & $(1/60,1/60,1/60)$ & $(1/45,1/90,1/90)$ & $(1/40,1/80,1/80)$ \\
Finite-difference order & $8$ & $8$ & $8$ \\
Runge--Kutta order & $4$ & $4$ & $4$ \\
Courant factor & $0.45$ & $0.45$ & $0.45$ \\
Kreiss--Oliger coefficient $\epsilon_{\rm KO}$ & $0.15$ & $0.15$ & $0.15$ \\
Interpolation orders $(x,t,\mathrm{patch})$ & $(5,2,5)$ & $(5,2,5)$ & $(5,2,5)$ \\
\colrule
\multicolumn{4}{c}{\textit{Horizon solve}} \\
AHFinderDirect zones per right angle $N$ & $54$ & $54$ & $54$ \\
Expansion norm tolerance & $10^{-8}$ & $10^{-8}$ & $10^{-8}$ \\
Maximum Newton iterations  & $400$ & $400$ & $400$ \\
$\Theta$-growth/nonshrink allowances & $(240,240)$ & $(240,240)$ & $(240,240)$ \\

\end{tabular}
\end{ruledtabular}
\end{table*}

\subsection{Horizon finding near formation}
\label{sec:horizon-finding}

The numerical MOTSs are represented as radial graphs; we relate this representation to the normal graphs used in Sec.~\ref{sec:theory}.  MOTS searches use \texttt{AHFinderDirect} \cite{Thornburg:2003sf}, which solves the nonlinear elliptic MOTS equation on each spatial slice and, like other widely used finders, requires the candidate surface to be a Strahlk{\"o}rper, meaning that it can be parametrized as $r=\bar{H}(\theta,\phi)$ about a given center.  Both common horizon branches are represented this way about the initial center of mass, and each is followed for as long as that representation stays nonsingular.  Let $\mathbf J$ denote the radial linearization of the MOTS equation. On a fixed slice, given an initial guess $\bar{H}^{(0)}$, the solver computes the first-order expansion of the MOTS equation at fixed $t$, 
\begin{equation}
    \Theta[\bar{H}^{(0)}+\delta \bar{H}]\approx\Theta[\bar{H}^{(0)}]+\mathbf{J}[\delta \bar{H}],
\end{equation}
and solves the linearized equation
\begin{equation}
    \mathbf J[\delta\bar H]
    =
    -\Theta[\bar H^{(0)}].
\end{equation}
The solver obtains $\delta\bar H$, then updates the candidate surface to $\bar{H}^{(1)} = \bar{H}^{(0)} + \delta \bar{H}$ and iterates until it converges to a surface that satisfies the MOTS equation.
At formation, the zero mode of the stability operator also makes the radial Jacobian $\mathbf J$ singular.

\begin{figure}[tbp]
    \centering
    \includegraphics[width=0.95\columnwidth]{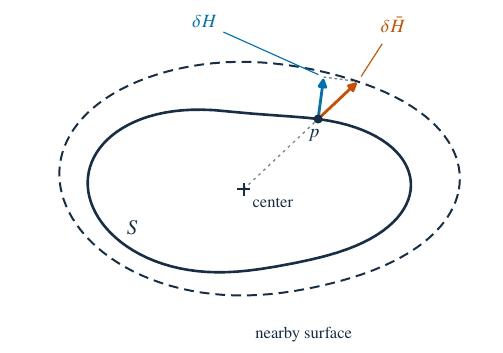}
    \caption{Radial and normal displacements of a surface. The solid curve denotes a reference surface, the dashed curve a nearby radial graph, and the cross the center of the radial coordinates. At $p$, the radial displacement associated with $\delta\bar H$ (orange) decomposes, to first order, into a normal displacement $\delta H$ (blue) and a tangential part (dotted).}
    \label{fig:radial-normal}
\end{figure}

At a reference surface, varying the normal graph height \(H\) of Sec.~\ref{representation} gives a first-order displacement along its outward unit normal. The Strahlk\"orper variable
\(\bar H\) is a coordinate radius.  Therefore a correction
\(\delta \bar H\) moves the surface along the coordinate radial ray at
fixed \((\theta,\phi)\).  This radial direction is not generally
normal to the surface (Fig.~\ref{fig:radial-normal}).

Only the normal part of this radial displacement changes the surface
geometrically to first order.  The tangential part only relabels points
on the same surface, and therefore does not contribute to the
linearized MOTS equation once \(\Theta_+=0\).  Thus the normal
deformation corresponding to a radial correction has the form
\begin{equation}
    \delta H = w\,\delta\bar H ,
    \label{eq:normal-radial}
\end{equation}
where
\begin{equation}
    w(\theta,\phi)=s_i\left(\frac{\partial x^i}{\partial r}\right)_{\theta,\phi}
\end{equation}
is the projection of the coordinate radial direction onto the outward
unit normal of the surface. For a valid radial representation, $w>0$, so multiplication by $w$ neither introduces nor removes a zero mode.

On a MOTS,
\begin{equation}
    \mathbf J[\delta\bar H]
    =
    L_\Sigma[w\,\delta\bar H],
\end{equation}
so the stability spectrum follows from
\begin{equation}
    \mathbf J\,\delta\bar H
    =
    \lambda\,w\,\delta\bar H,
\end{equation}
whose principal eigenvalue $\lambda_0$ is the one with smallest real part.  

\subsection{Formation surface and branch fields}
\label{sec:branch-fields}

For a horizon quantity $\mathcal{Q}$ evaluated on both branches, define 
\begin{eqnarray}
    \mathbb E[\mathcal{Q}]&=&(\mathcal{Q}_++\mathcal{Q}_-)/2\\
    \mathbb O[\mathcal{Q}]&=&(\mathcal{Q}_+-\mathcal{Q}_-)/2,
\end{eqnarray}
and similarly for $\bar{H}(\theta,\phi)$, applied pointwise at each $(\theta,\phi)$.
Let $t_1$ be the first slice on which both common horizons are found numerically.  We
reconstruct the common horizon geometry from the first five pairs, which span $t-t_*\lesssim0.03M$.  Because \texttt{AHFinderDirect} represents both branches on the
same angular grid about the same origin, half their radial separation at each $(\theta,\phi)$ is $\mathbb O[\bar H]$.
The radial and normal representations are related at leading order by the positive factor $w$ of Eq.~\eqref{eq:normal-radial}, so $\mathbb O[\bar H]=w^{-1}_*\mathcal B H_0\,(\delta t)^{1/2}+O((\delta t)^{3/2})$ carries only half-integer powers of $\delta t$, and its square only integer ones.  The squared separation therefore has a quadratic expansion through next order,
\begin{equation}
    \frac{1}{4\pi}\int\mathbb O[\bar H(t)]^2\,d\Omega
    =\kappa_1(t-t_*)+\kappa_2(t-t_*)^2+O\!\left((t-t_*)^3\right),
    \label{eq:radial-separation-fit}
\end{equation}
with $d\Omega$ the coordinate solid angle.  The left-hand side is a mean square that vanishes at formation and grows after it, so $\kappa_1>0$.  Dropping $\kappa_2$ gives the leading-order estimate.  Including the quadratic term substantially reduces the dependence on the fit window (Appendix~\ref{app:fit-window}), so we use the formation time inferred from the quadratic fit to reconstruct $S_*$.  

The radial graph has the same branch structure,
\begin{equation}
    \bar H_\pm(t)=\bar H_*\pm\mathcal A_{\bar H}(\delta t)^{1/2}
    +\mathcal B_{\bar H}\,\delta t+O((\delta t)^{3/2}),
    \label{eq:branch-series}
\end{equation}
and every half-integer power in it flips sign between the branches, so
$\mathbb E$ and $\mathbb O$ separate them,
\begin{align}
    \mathbb E[\bar H]&=\bar H_*+\mathcal B_{\bar H}\,\delta t+O((\delta t)^2),
    \nonumber\\
    \mathbb O[\bar H]&=\mathcal A_{\bar H}(\delta t)^{1/2}+O((\delta t)^{3/2}).
    \label{eq:parity-series}
\end{align}
$\mathbb E[\bar H]$ is an ordinary Taylor series in $\delta t$ whose constant term is $\bar H_*$, so we fit
\begin{equation}
    \mathbb E[\bar H]
    =\bar H_*+\mathcal B_{\bar H}\,\delta t+\mathcal D_{\bar H}\,(\delta t)^2
    +O\!\left((\delta t)^3\right)
    \label{eq:mean-quadratic-fit}
\end{equation}
and read $\bar H_*$ off at $\delta t=0$.  

We then reconstruct the metric at $t_*$ by linear componentwise interpolation between the data on the two stored slices $t_a<t_*<t_b$ that bracket it,
\begin{equation}
    \gamma_{ij}(t_*)
    =\frac{(t_b-t_*)\,\gamma_{ij}(t_a)+(t_*-t_a)\,\gamma_{ij}(t_b)}{t_b-t_a},
    \label{eq:metric-interpolation}
\end{equation}
and apply the same interpolation to the extrinsic curvature $K_{ij}$.  We compute the outward unit normal $s_*^i$ of
Eq.~\eqref{eq:star-normal}, the expansion residual and the stability eigenvalue there,
to confirm that we recover the surface on which the common horizon first forms.  Table~\ref{tab:formation-surface} reports the expansion residuals and stability eigenvalues of the reconstructed surface. 
\begin{table}[htbp]
\caption{Verification of the formation surface.  Using the metric and extrinsic curvature interpolated to $t_*$, we evaluate the outgoing expansion and stability spectrum on $S_*$. The expansion residuals and near-zero principal eigenvalue quantify how closely the reconstruction satisfies $\Theta_+=0$ and $\lambda_0=0$. The last row gives $\lambda_1$, defined by the next smallest real part; complex-conjugate pairs are shown together.}
\label{tab:formation-surface}
\begin{ruledtabular}
\begin{tabular}{lccc}
 & Sim.~I & Sim.~II & Sim.~III \\
\colrule
RMS $\Theta_+[S_*]$ $(M^{-1})$ & $5.6\times10^{-5}$ & $3.2\times10^{-4}$ & $1.1\times10^{-4}$ \\
$\max\bigl|\Theta_+[S_*]\bigr|$ $(M^{-1})$ & $1.4\times10^{-3}$ & $8.8\times10^{-3}$ & $3.8\times10^{-3}$ \\
$\lambda_0[S_*]$ $(M^{-2})$ & $-1.8\times10^{-6}$ & $-6.8\times10^{-6}$ & $7.7\times10^{-6}$ \\
$\lambda_1[S_*]$ $(M^{-2})$ & $0.70\pm0.17i$ & $0.64\pm0.08i$ & $0.65$ \\
\end{tabular}
\end{ruledtabular}
\end{table}

We now convert each radial representation into the outward normal representation with respect to $S_*$. For each $p\in S_*$ we solve
\begin{equation}
    \Bigl.\bigl(r-\bar H_\pm(\theta,\phi;t)\bigr)\Bigr|_{x^i=x_*^i(p)+\delta H_\pm s_*^i(p)}=0
    \label{eq:normal-ray-intersection}
\end{equation}
for $\delta H_\pm$.

The branch average determines the common drift, while half the branch separation determines the opening field:

\begin{align}
    \mathbb E[\delta H(\delta t)]
    &=\delta t\,H_d+O((\delta t)^{2}),
    \nonumber\\
    \mathbb O[\delta H(\delta t)]
    &=\mathcal B(\delta t)^{1/2}H_0+O((\delta t)^{3/2}).
\end{align}

The $\mathbb O$ fit estimates the principal zero mode $H_0$ of the stability operator on $S_*$, with exact identification in the limit $\delta t\to0^+$; we fix its normalization by $\int_{S_*}H_0^2dA_*=1$.  Recombining the two equations above gives
\begin{equation}
    \delta H_\pm(\delta t)
    =\pm \mathcal B(\delta t)^{1/2}H_0
    +\delta t\,H_d
    +O((\delta t)^{3/2}).
    \label{eq:n54-shape-fit}
\end{equation}

\begin{figure*}[t]
  \centering
  \includegraphics[width=0.98\textwidth]{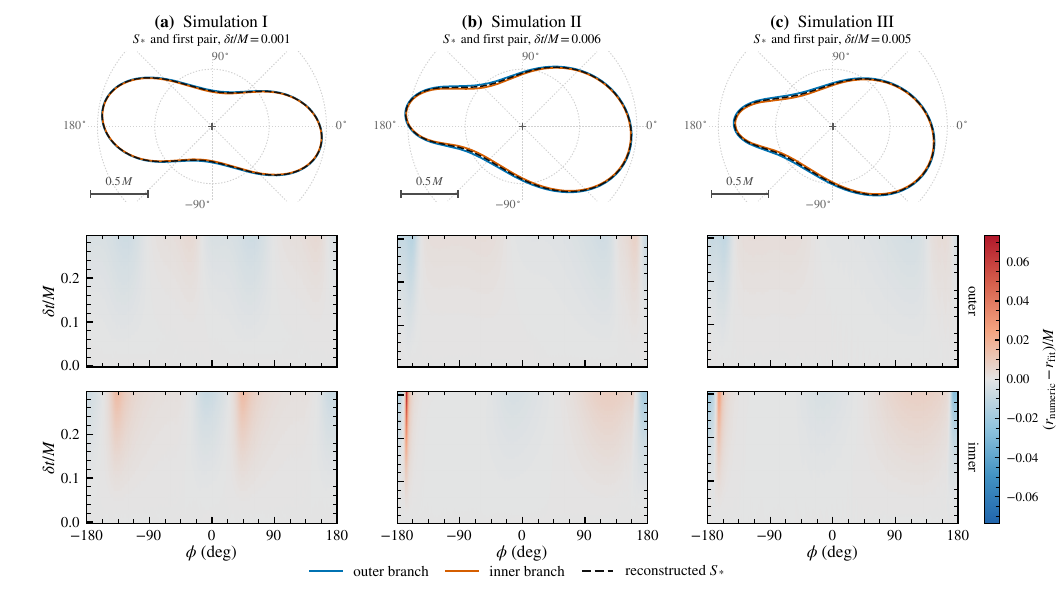}
  \caption{Cross sections of the geometric reconstruction of the local MOTT for (a)--(c) Simulations~I--III. Each section lies in the plane orthogonal to the outer branch's coordinate spin axis $z'$ at $t_1$. The in-plane orientation is set by the puncture-separation vector from \texttt{PunctureTracker} \cite{Loffler:2011ay}, projected into this plane. For the precessing Simulation~III, the plane does not coincide with the coordinate equator.  Top row: the reconstructed formation surface $S_*$ (black dashed) against the first numerical pair, on a polar graticule about the Strahlk\"orper origin with the bar giving the scale.  Lower rows: the residual $r_{\text{numeric}}-r_{\text{fit}}$ left by Eq.~\eqref{eq:n54-shape-fit} on that section, for the outer and inner branches. The residuals are expressed in radial coordinates. The prediction is fitted to the first five pairs, $\delta t\lesssim0.03M$, while the maps extend to $\delta t\simeq0.3M$. Residuals are small near formation and increase at later times. }
  \label{fig:n54-geometry-comparison}
\end{figure*}

Figure~\ref{fig:n54-geometry-comparison} compares the reconstructed MOTT across the three simulations. The maps run to $\delta t\simeq0.3M$, an order of magnitude beyond the window in which Eq.~\eqref{eq:n54-shape-fit} was fitted.

\begin{figure*}[t]
  \centering
  \includegraphics[width=0.985\textwidth]{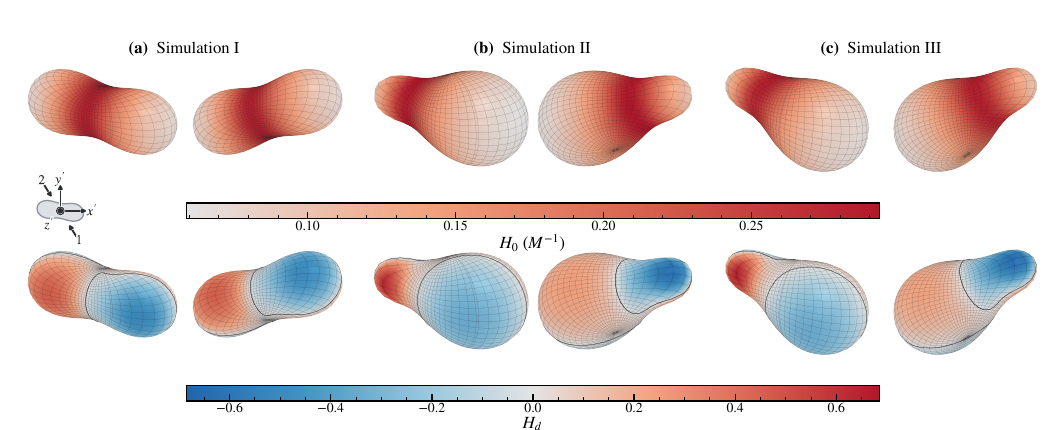}
  \caption{Deformation fields for (a)--(c) Simulations~I--III, on the reconstructed $S_*$.  The upper row shows the normalized principal zero mode $H_0$ estimated from the branch separation, which controls the separation of the outer and inner common MOTSs. The lower row shows $H_d$. Within each simulation the two panels show opposite viewing directions of the same surface.  The key beside the upper color bar shows $S_*$ sectioned on the plane orthogonal to $z'$; $x'$ is the direction of puncture separation projected into this plane and $z'$ points out of the page.  The numbered arrows show the camera lines of sight, $20^\circ$ above and below that plane; the numbers indicate their left-to-right order within each pair.  A common coordinate scale is used for all renderings; each field has one color scale shared across the three simulations, and black curves mark $H_d=0$.}
  \label{fig:n54-surface-fields}
\end{figure*}

Figure~\ref{fig:n54-surface-fields} shows the resulting $H_0$ and $H_d$.  The fitted $H_0$ is positive everywhere in all three simulations, as predicted by the stability analysis, and is largest in the neck region.  The field $H_d$ changes sign across the surface and carries the displacement and deformation shared by the two branches.

\subsection{Scaling of horizon quantities}
\label{sec:scaling-fits}

\begin{figure*}[t]
  \centering
  \includegraphics[width=\textwidth]{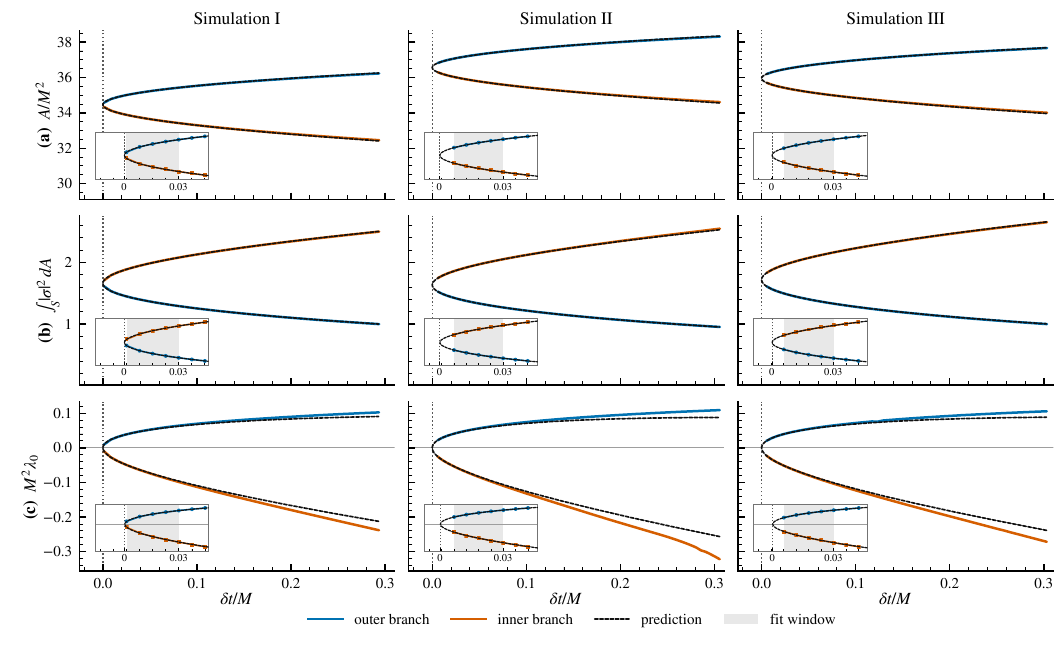}
  \caption{Fits to the predicted scaling laws for Simulations~I--III (columns).  Rows (a)--(c) show $A/M^2$, $\int_S|\sigma|^2dA$, and $M^2\lambda_0$.  Blue and vermilion denote the outer and inner branches.  Dashed black curves show fits to Eq.~\eqref{eq:Q-order-by-order} for the area and integrated squared shear, and to Eq.~\eqref{eq:lambda0-slope} for the principal eigenvalue, and gray bands in the insets mark the five-pair fit windows.  The insets magnify the region near formation, and the inset ticks mark the cadence of horizon searches.}
  \label{fig:n54-quasilocal-scaling}
\end{figure*}

We test Eq.~\eqref{eq:Q-order-by-order} with the area $A=\int_SdA$ and the integrated squared magnitude of the outgoing shear $\int_S|\sigma|^2dA$, and compare the result with the principal stability eigenvalue $\lambda_0$.  The shear is a complex scalar field computed by \texttt{QuasiLocalMeasures} \cite{Dreyer:2003isolatedNR,SchnetterKrishnanBeyer:2006dynamicalNR} using the Newman--Penrose formalism. With the outgoing null normal $\ell^a$ normalized as in Sec.~\ref{sec:basic}, its definition is
\begin{equation}
    \sigma=-m^a m^b\nabla_b\ell_a,
    \qquad m^a m_a=0,\quad m^a\bar m_a=1,
    \label{eq:complex-shear-definition}
\end{equation}
where $m^a$ is a complex tangent vector on the MOTS and $\nabla_a$ is the spacetime covariant derivative. We integrate $|\sigma|^2$ over the MOTS two-sphere using the induced metric, also computed by \texttt{QuasiLocalMeasures}.  For each quantity in each simulation, the same form is fitted to the first five branch pairs, with $t_{*,\mathcal Q}$ fitted separately for each quantity in place of $t_*$.

Because the exponents are held at their predicted values, these fits test whether the formation times agree.  The lower panel of Table~\ref{tab:formation-times} reports fits with the exponent free, using
\begin{align}
    \bigl|\mathbb O[\mathcal Q]\bigr|
    &=\mathcal A_{\mathcal Q}\,(t-t_{*,\mathcal Q})^{p}
    \nonumber\\
    \text{and}\quad
    \bigl|\mathbb O[\mathcal Q]\bigr|
    &=(t-t_{*,\mathcal Q})^{p}
    \bigl[\mathcal A_{\mathcal Q}+\mathcal C_{\mathcal Q}\,(t-t_{*,\mathcal Q})\bigr]
    \label{eq:free-exponent-models}
\end{align}
at leading and next order, with $p$, $t_{*,\mathcal Q}$, and the amplitudes all free.  Figure~\ref{fig:n54-quasilocal-scaling} shows the fits continued beyond the shaded five-pair intervals.

The principal eigenvalue is not a surface functional of the form assumed in Eq.~\eqref{eq:Q-order-by-order}, and its branch expansion, Eq.~\eqref{eq:lambda0-slope}, was obtained separately in Sec.~\ref{sec:eigenvalue-expansion}.  It is fitted here in the same way.

Table~\ref{tab:formation-times} reports the interval between the inferred formation time and the first detected pair, together with the exponents obtained from the five-pair fits, while Appendix~\ref{app:fit-window} shows their dependence on windows of four through fourteen branch pairs.  Across that scan, the three surface estimators remain consistent.

The leading-order model shows larger deviations for $\lambda_0$ than for the two surface functionals.  Its estimate of formation time differs from the other estimates by about $0.25\times10^{-3}M$, and its fitted exponent exceeds $1/2$. Including the next-order term improves agreement among the estimates of formation time and reduces the dependence of the $\lambda_0$ estimate on the fit window by a factor of five or more (Fig.~\ref{fig:fit-window}). Table~\ref{tab:formation-times} gives the corresponding exponents, including the larger next-order deviation for $\lambda_0$ in Simulation~II.

\begin{table*}[t]
\caption{Estimates of formation time and exponent from fits to the first five pairs. The upper panel reports the interval $t_1-t_{*,\mathcal Q}$ between the inferred formation time and the first detected pair, and the lower panel reports the fitted exponent.  For geometry, we fit the angular mean of $\mathbb O[\bar H]^2$ in Eq.~\eqref{eq:radial-separation-fit}, and for the other three quantities we fit $\mathbb O[\mathcal Q]^2$ in the same way. The exponents are refitted from $\bigl|\mathbb O[\mathcal Q]\bigr|$ for the latter and from the root-mean-square of $\mathbb O[\bar H]$ for geometry.  Dependence on the fit window is shown in Appendix~\ref{app:fit-window}. Over windows of four to eight pairs, the next-order times change by at most $1.1\times10^{-5}M$ for the three surface quantities and $6.9\times10^{-5}M$ for $\lambda_0$, and the next-order exponents by at most $1.0\times10^{-3}$ and $3.4\times10^{-2}$.  The left block uses the leading-order model and the right block includes the next-order term.  At leading order the three surface
quantities agree and $\lambda_0$ stands apart in both panels; including that term brings the eigenvalue estimates closer to the others in both panels.}
\label{tab:formation-times}
\begin{ruledtabular}
\begin{tabular}{lcccccc}
 & \multicolumn{3}{c}{$O(\delta t)$} & \multicolumn{3}{c}{$O((\delta t)^{3/2})$} \\
\cline{2-4}\cline{5-7}
Estimator & Sim.~I & Sim.~II & Sim.~III & Sim.~I & Sim.~II & Sim.~III \\
\colrule
\multicolumn{7}{c}{\textit{Time from formation to the first detected pair $t_1-t_{*,\mathcal Q}$, in units of $10^{-3}M$}} \\
Geometry & $1.18$ & $5.77$ & $5.25$ & $1.14$ & $5.75$ & $5.215$ \\
Area $A$ & $1.18$ & $5.76$ & $5.25$ & $1.14$ & $5.75$ & $5.216$ \\
Integrated squared shear & $1.16$ & $5.74$ & $5.22$ & $1.14$ & $5.75$ & $5.214$ \\
Principal eigenvalue $\lambda_0$ & $0.95$ & $5.52$ & $4.97$ & $1.15$ & $5.80$ & $5.19$ \\
\colrule
\multicolumn{7}{c}{\textit{Exponent $p$, refitted with $p$ free}} \\
Geometry & $0.4975$ & $0.4989$ & $0.4975$ & $0.4995$ & $0.5018$ & $0.4996$ \\
Area $A$ & $0.4972$ & $0.4990$ & $0.4978$ & $0.4995$ & $0.5017$ & $0.4995$ \\
Integrated squared shear & $0.4987$ & $0.5007$ & $0.4995$ & $0.4995$ & $0.5020$ & $0.4998$ \\
Principal eigenvalue $\lambda_0$ & $0.5146$ & $0.5188$ & $0.5155$ & $0.4977$ & $0.4890$ & $0.5024$ \\
\end{tabular}
\end{ruledtabular}
\end{table*}

\section{Discussion and outlook}\label{sec:discussion}

We derive a local expansion of common horizon formation that gives square-root branch separation with a shared linear drift. Past isolation forces a zero mode of the MOTS stability operator, and outermost stability identifies it as the positive principal mode $H_0$ and makes the kernel of $L_\Sigma$ one-dimensional. Lyapunov--Schmidt reduction determines the amplitude of the zero mode through a scalar equation, while the complementary equation fixes the remaining deformation. The resulting MOTT is tangent to the formation slice at $S_*$, where slice time has a quadratic minimum along the tube. Retaining the linear drift gives the tilted parabola described by Eq.~\eqref{eq:Hexpansion}, which specifies the surface displacement through linear order in time. Nearby later slices intersect the smooth tube in an outer and an inner common MOTS whose separation grows as $(t-t_*)^{1/2}$.  The exponent is universal, and the coefficients are not phenomenological. Each is computable from geometric data on the formation surface.  All three simulations exhibit the predicted geometry, opening and drift fields, and scaling of horizon quantities. Their separately fitted formation times agree (Table~\ref{tab:formation-times}), and the reconstructed $S_*$ satisfies $\Theta_+=0$ and $\lambda_0=0$ to the accuracy reported in Table~\ref{tab:formation-surface}.  The analysis assumes no symmetry, and the eccentric, precessing, unequal-mass Simulation~III also exhibits the predicted branch structure.

The branch expansion, Eq.~\eqref{eq:Hexpansion}, describes the surface displacement through two fields with distinct physical roles.  The zero-mode contribution $\pm\mathcal B(\delta t)^{1/2}H_0$ is associated with pair creation: since $H_0$ is strictly positive, the outer branch lies outside the inner branch over the whole surface. In all three simulations, the measured $H_0$ is indeed positive.  The common drift $\delta t\,H_d$ changes sign across the surface and contains the coordinate motion and deformation shared by both branches. Resolving this field into the normal components of rigid coordinate translation and rotation would provide quantities to compare with independently measured recoil and angular momentum \cite{IozzoKheraEtAl:2021remnantComparison}. We leave this decomposition and comparison to future work.  

For a fixed foliation, consider a smooth, future-oriented relabeling $\tilde t=f(t)$ with $f'(t_*)>0$, and set $\delta\tilde t=\tilde t-f(t_*)$. Since $\delta\tilde t=f'(t_*)\delta t+O((\delta t)^2)$, the branch expansion becomes
\begin{equation}
\begin{aligned}
    \mathcal Q_\pm-\mathcal Q_*
    ={}&\pm\frac{\mathcal A_{\mathcal Q}}{\sqrt{f'(t_*)}}
        (\delta\tilde t)^{1/2}\\
    &+\frac{\mathcal B_{\mathcal Q}}{f'(t_*)}\,\delta\tilde t
      +O((\delta\tilde t)^{3/2}).
\end{aligned}
\label{eq:time-relabeling}
\end{equation}
Thus the exponent and the opposite signs of the opening term are unchanged, while both amplitudes rescale and the linear term remains shared. This relabeling leaves the slices themselves unchanged. The coefficient $\mathcal A_{\mathcal Q}$ carries the units of $\mathcal Q$ times $(\mathrm{time})^{-1/2}$; comparisons across simulations therefore require consistent time and shape conventions.  The divergent growth rate with respect to slice time has the same geometric origin. At the tangency, $dt/d\tau$ vanishes, so $dA/dt$ can diverge even though the area varies smoothly with the worldtube parameter $\tau$. 

The same local analysis, applied about later tangencies, also provides a model for subsequent changes in merger horizon structure. Numerical evolutions show MOTSs appearing and disappearing in pairs, worldtubes turning in slice time, and leaves developing self-intersections \cite{Pook-Kolb:2019iao,Pook-Kolb:2019ssg,Pook-Kolb:2021gsh,Booth:2021sow,Pook-Kolb:2021jpd,Kastha:2026cusp}. Related bifurcation and annihilation sequences occur among interior MOTSs in exact spherically symmetric spacetimes \cite{HennigarEtAl:2022interiorMOTS}. The reduction extends to an isolated zero eigenvalue with a one-dimensional kernel when the same two nondegeneracy conditions hold. At such a tangency, the same square-root law follows.  The principal mode is distinguished by its positivity, which makes the two branches nest.  At interior tangencies the vanishing eigenvalue can be nonprincipal, and the branches need not nest.  A disappearance is the fold run backward in time, with the separation closing as $(t_{\rm tan}-t)^{1/2}$ toward the tangency time $t_{\rm tan}$ and the crossing eigenvalue passing through zero with opposite square-root slopes on the two branches.  The first common-horizon formation studied here is one instance of this mechanism. The concurrent work of Booth, Cox, and Okpala~\cite{BoothCoxOkpala:2026bifurcation} studies related bifurcation structure in horizon evolutions.  Interior surfaces need not be stable. On segments where the stability operator remains invertible, the tube continues smoothly; at each nondegenerate tangency we have a computable local expansion.  Gluing these local descriptions along a worldtube may provide an analytic description of the merger's interior horizon structure.  

The inner branch connects the formation geometry to the subsequent evolution of horizon area. Axisymmetric studies show the inner common MOTS approaching the individual horizons and developing cusps at contact, with its area approaching the sum of the individual areas \cite{Pook-Kolb:2019iao,Kastha:2026cusp}. The formation expansion and the late contact behavior constrain different portions of this evolution. Both common branches begin with area $A_*$, and the theory determines the leading separation of their areas. The inner area subsequently approaches the sum of the progenitor areas, while the outer area approaches the final remnant horizon area, which can be inferred from gravitational-wave estimates of the remnant mass and spin. Extending this comparison to nonaxisymmetric mergers would test how $A_*$ relates to $A_1+A_2$ and how far the formation expansion describes the inner branch before contact asymptotics become relevant. The standard flux laws for dynamical horizons \cite{AshtekarKrishnan:2002flux,AshtekarKrishnan:2003properties} describe area growth along a smooth dynamical horizon; by themselves, they do not relate the disconnected progenitor horizons to the newly formed common horizon. 

A second application is gravitational-wave tomography: inferring horizon dynamics from radiation at null infinity. In a perturbative ringdown setting, explicit relations between the horizon and asymptotic radiative fields establish this connection analytically \cite{RibesMetidieriBongaKrishnan:2025tomography}. The outgoing horizon shear $\sigma$ offers a way to extend this description toward merger. Numerical simulations find correlations between the shear and the gravitational-wave news on the individual horizons during inspiral \cite{Prasad:2020news} and on the outer common horizon after formation \cite{Prasad:2025shear}.

At formation, the apparent-horizon boundary jumps outward while the individual horizons persist, whereas the outgoing waveform remains continuous. The common horizon forms inside the event horizon, and its birth sends no signal to null infinity, so the correlation reflects the spacetime geometry that determines both the horizon and the radiation. Our local formation expansion provides the geometry needed to calculate the shear near $t_*$, including both the square-root branch separation and the shared linear drift. A tomographic extension to the merger regime would relate the formation data $(t_*,\,\mathcal B,\,H_0,\,H_d)$ to the outgoing radiation. Comparing the resulting shear with the news would test whether common horizon formation has an identifiable counterpart in the gravitational waveform. Conversely, what the formation expansion leaves out is also informative. It is purely geometric, since it follows from how the slices cut a smooth MOTT, and it is unlikely to contain the quasinormal ringing of the remnant \cite{Mourier:2021commonHorizonQNMs,Chen:2022commonHorizonMultipoles,Prasad:2025shear}. The time at which the outer common horizon departs from Eq.~\eqref{eq:Hexpansion} and begins to ring down would then indicate when the remnant can be described by black hole perturbation theory.

The local expansion explains the loss of conditioning in horizon searches using Newton's method and provides a way to reconstruct formation from later surfaces.  Standard finders \cite{Thornburg:2003sf} linearize the elliptic MOTS equation about a candidate surface and solve for a correction. At formation, the zero mode makes the Jacobian $\mathbf J$ of Sec.~\ref{sec:horizon-finding} singular, so the ordinary Newton step based on its inverse is no longer defined. Near formation, poor conditioning can delay detection of the first pair.  In Simulation~II, the interval from the fitted formation time to the first detected pair, $t_1-t_*=5.75\times10^{-3}M$, exceeds the $5.0\times10^{-3}M$ search interval: at our settings (Table~\ref{tab:nr-setup}), the finder did not converge on a slice where the reconstructed branch structure places both MOTSs. Fits to later, converged pairs recover $t_*$ and $S_*$, whose expansion and stability spectrum are evaluated in Table~\ref{tab:formation-surface}.

These reconstructed surfaces could seed searches near formation, and monitoring the approach to a zero mode could guide the choice of search times.

\section*{Data Availability}
The processed data supporting the findings of this study, together with provenance information for the raw simulation outputs, are available from the corresponding author upon reasonable request.  The raw numerical relativity outputs are not available because of their size.

\begin{acknowledgments}
The U.S. National Science Foundation (NSF) supports YCX under Grant No.~\mbox{PHY-2608011} and VP under Grant Nos.~\mbox{PHY-2608388}, \mbox{AST-2307147}, \mbox{AST-2606673}, \mbox{PHY-2608011}, \mbox{PHY-2308886}, and \mbox{PHY-2309064}.
YCX used Anvil at Purdue University \cite{anvil} and Bridges-2 at the Pittsburgh Supercomputing Center \cite{bridges2} through allocations PHY260122, PHY260295, and PHY250294 from the Advanced Cyberinfrastructure Coordination Ecosystem: Services \& Support (ACCESS) program \cite{access}, which is supported by U.S. National Science Foundation grants \#2138259, \#2138286, \#2138307, \#2137603, and \#2138296.
YCX thanks Abhay Ashtekar, Ivan Booth, Graham Cox, Peter Hintz, and B.~S.~Sathyaprakash for helpful discussions.
The authors used GPT-6 Astra, GPT-5.6 Sol, Opus 5, and Fable 5 for assistance with part of the coding and with the preparation of illustrative figures.
\end{acknowledgments}
\appendix
\section{Convergence tests}
We assess resolution dependence using $n=28$, $32$, and $36$ cells across the finest-grid radius for Simulations~I and II.  Simulation~III evolves the full domain without equatorial symmetry, and a three-resolution series for it was beyond the available computational resources; its production run uses $n=32$.  Convergence in the wave zone is assessed from the mismatch of the extracted $\Psi_4$ at the largest extraction sphere, while convergence on the common horizon is assessed from the mismatch of the Newman--Penrose shear $\sigma$ near its north pole.
\begin{figure}[htbp]
  \centering
  \includegraphics[width=\columnwidth]{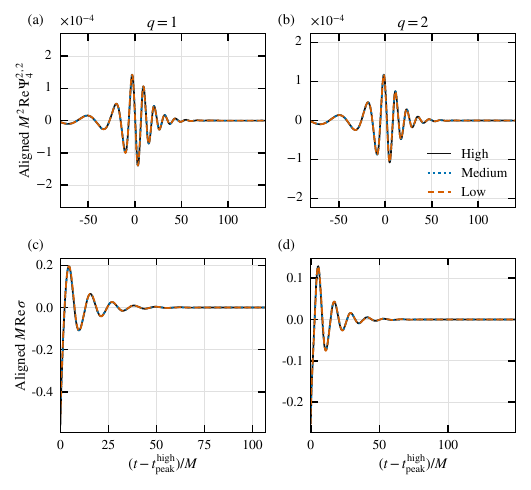}
  \caption{Convergence of the wave-zone and horizon diagnostics.  Panels (a) and (b) show the dominant $(\ell,m)=(2,2)$ mode of $\Psi_4$ at extraction radius $r=500M$; panels (c) and (d) show the Newman--Penrose shear $\sigma$ of the outgoing null normal near the north pole of the common horizon.  The left and right columns show Simulations~I and II, respectively.  Each series is aligned to the highest resolution by the relative time shift and overall complex rescaling that maximize the overlap; the resulting mismatches are listed in Table~\ref{tab:convergence-mismatch}.}
  \label{fig:convergence}
\end{figure}

\begin{table}[htbp]
\caption{Mismatch $\mathcal M$ between successive resolutions.  The $\Psi_4$ mismatch uses the waveform reconstructed from all modes with $2\leq\ell\leq8$ at extraction radius $r=500M$; the shear mismatch uses $\sigma$ near the north pole of the common horizon.  Both are minimized over relative time and phase shifts and integrated over the full overlapping interval.}
\label{tab:convergence-mismatch}
\begin{ruledtabular}
\begin{tabular}{llcc}
Diagnostic & Resolution pair & Sim.~I & Sim.~II \\
\colrule
\multirow{2}{*}{$\Psi_4$} & low--medium  & $4.53\times10^{-3}$ & $4.16\times10^{-3}$ \\
               & medium--high & $1.07\times10^{-3}$ & $3.00\times10^{-3}$ \\
\multirow{2}{*}{$\sigma$}       & low--medium  & $1.56\times10^{-3}$ & $4.24\times10^{-4}$ \\
               & medium--high & $2.09\times10^{-5}$ & $8.94\times10^{-5}$ \\
\end{tabular}
\end{ruledtabular}
\end{table}

Both mismatches decrease with increasing resolution, as listed in Table~\ref{tab:convergence-mismatch}.

\section{Dependence on the fit window}
\label{app:fit-window}

\begin{figure*}[t]
  \centering
  \includegraphics[width=\textwidth]{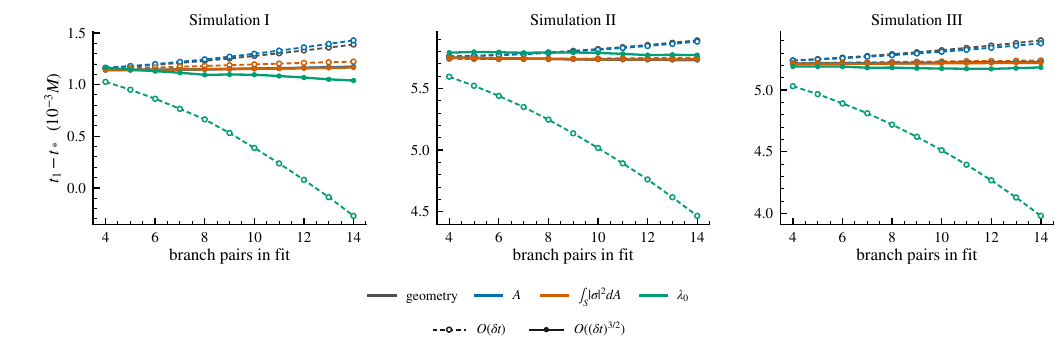}
  \caption{Time between formation and the first detected pair as a function of the number of fitted branch pairs, for
  Simulations~I--III, for each of the four estimators of Table~\ref{tab:formation-times}.
  For geometry, we fit the angular mean of $\mathbb O[\bar H]^2$ in Eq.~\eqref{eq:radial-separation-fit}, and for the other three quantities we fit $\mathbb O[\mathcal Q]^2$. The next-order estimates vary less with the fit window than the leading-order estimates.}
  \label{fig:fit-window}
\end{figure*}

Table~\ref{tab:formation-times} uses a fit window containing five pairs. In Fig.~\ref{fig:fit-window}, we vary the window from four to fourteen branch pairs. The leading-order estimates depend systematically on the window, most strongly for $\lambda_0$. For sufficiently long windows, the leading-order eigenvalue fit places $t_*$ later than the first detected pair, indicating breakdown of that truncated model. Including the next-order term reduces the window dependence, and the estimates from geometry, area, and integrated squared shear agree closely across the range shown.

\bibliography{refs}
\end{document}